\documentclass[a4paper,usenatbib,fleqn]{mn2e}

\usepackage[T1]{fontenc}
\usepackage{ae,aecompl}

\usepackage[pdftex]{graphicx}	% Including figure files
\usepackage{amsmath}	% Advanced maths commands
\usepackage{amssymb}	% Extra maths symbols
\usepackage{color}      % To show changes in the manuscript for
\IfFileExists{url.sty}{\usepackage{url}}
                      {\newcommand{\url}{\texttt}}

\pdfpageheight\paperheight
\pdfpagewidth\paperwidth

\providecommand{\tabularnewline}{\\}
\newcommand{\kmps}{\mathrm{km~s^{-1}}}
\newcommand{\Kelvin}{\mathrm{K}}
\newcommand{\Msun}{\mathrm{M_{\sun}}}
\newcommand{\Rsun}{\mathrm{R_{\sun}}}
\newcommand{\Lsun}{\mathrm{L_{\sun}}}
\newcommand{\MsunPerYear}{\mathrm{M_{\sun}\,yr^{-1}}}

\newcommand{\ion}[2]{#1$\,${\sc {#2}}}   % ion, i.e., CII = \ion{C}{ii}

\title[Wind-launching regions of Herbig Be star]{Probing
  the wind launching regions of the Herbig Be star HD\,58647 with
  high spectral resolution interferometry\thanks{Based on
      observations made with ESO Telescopes at the La Silla Paranal
      Observatory under programmes ID~094.C-0164, 086.C-0267 and 084.C-0170.}}  

\author[R. Kurosawa et\,al.]{Ryuichi Kurosawa$^{1}$\thanks{E-mail:ryuichi.kurosawa@gmail.com}, 
A.~Kreplin$^{2}$, G.~Weigelt$^{1}$,  A.~Natta$^{3,4}$,  M.~Benisty$^{5}$, \newauthor 
Andrea Isella$^{6}$, Eric Tatulli$^{5}$, F.~Massi$^{3}$, Leonardo
Testi$^{3,7,8}$,  Stefan Kraus$^{2}$, \newauthor
G.~Duvert$^{5}$,  Romain G.~Petrov$^{9}$ and Ph.~Stee$^{9}$\\
$^{1}$Max-Planck-Institut f\"{u}r Radioastronomie, Auf dem H\"{u}gel 69, 53121 Bonn, Germany\\
$^{2}$School of Physics, University of Exeter, Stocker Road, Exeter, EX4~4QL, United Kingdom\\
$^{3}$INAF/Osservatorio Astrofisico of Arcetri, Largo E. Fermi, 5, 50125 Firenze, Italy\\
$^{4}$Dublin Institute for Advanced Studies, School of Cosmic Physics, 31 Fitzwilliam Place, Dublin 2, Ireland\\
$^{5}$Univ. Grenoble Alpes, IPAG, F-38000 Grenoble, France;  CNRS, IPAG, F-38000 Grenoble, France\\
$^{6}$Department of Physics and Astronomy, Rice University, 6100 Main Street, Houston, TX 77005, USA\\
$^{7}$European Southern Observatory (ESO) Headquarters, Karl-Schwarzschild-Str. 2, D-85748 Garching bei M\"{u}nchen, Germany\\
$^{8}$Excellence Cluster Universe, Boltzmannstr. 2, D-85748 Garching bei M\"{u}nchen, Germany\\
$^{9}$Laboratoire Lagrange, Universit\'e C\^ote d'Azur, Observatoire de la C\^ote d'Azur, CNRS, Boulevard de l'Observatoire, CS 34229, \\
06304 Nice cedex 4, France\\
}

\date{Accepted XXX. Received YYY; in original form ZZZ} 
\pubyear{2015}

\begin{document}
% Don't change these lines
\label{firstpage} 
\pagerange{\pageref{firstpage}--\pageref{lastpage}}
\maketitle

% Abstract of the paper
\begin{abstract}
We present a study of the wind launching region of the Herbig Be star
HD~58647 using high angular ($\lambda/2B=0''.003$) and high spectral ($R=12000$) resolution
interferometric VLTI-AMBER observations of the near-infrared hydrogen emission
line, Br$\gamma$.  The star displays double peaks in both
Br$\gamma$ line profile and wavelength-dependent visibilities. The
wavelength-dependent differential phases show 
S-shaped variations around the line centre. The visibility level
increases in the line (by $\sim 0.1$) at the longest projected baseline
(88\,m), indicating that the size of the line emission region is
smaller than the size of the K-band continuum-emitting region,
which is expected to arise near the dust sublimation radius of the
accretion disc. The data have been analysed using radiative transfer
models to probe the geometry, size and physical properties of the wind
that is emitting Br$\gamma$.  We find that a model with a small
magnetosphere and a disc wind with its inner radius located just
outside of the magnetosphere can well reproduce the observed
Br$\gamma$ profile, wavelength-dependent visibilities, differential
and closure phases, simultaneously. The mass-accretion and mass-loss
rates adopted for the model are $\dot{M}_{\mathrm{a}}=3.5\times10^{-7}$ and
$\dot{M}_{\mathrm{dw}}=4.5\times10^{-8}\,\MsunPerYear$, respectively 
($\dot{M}_{\mathrm{dw}}/\dot{M}_{\mathrm{a}}=0.13$).  
Consequently, about 60~per~cent of the angular momentum loss
rate required for a steady accretion with the measured
accretion rate is provide by the disc wind.
The small magnetosphere in HD~58647 does not contribute to the
Br$\gamma$ line emission significantly.

\end{abstract}


\section{Introduction}

\label{sec:intro}

%%%%%%%%%%%%%%%%%%%%%%%%%%%%%%%%%%%%%%%%%%%%%%%%%%%%%%%%%
\begin{table*}

\caption{Summary of known properties of HD\,58647. \label{tab:published-parameters}}

\begin{center}

\begin{tabular}{cccccccccccc}
\hline 
$M_{*}$ & $R_{*}$ & $L_{*}$ & Sp. Type & Age & $T_{\mathrm{eff}}$ & $v\sin i$ & $\dot{M_{a}}$ & d & K & H & Av\tabularnewline
($\Msun$) & ($\Rsun$) & ($\Lsun$) & \ldots{} & (Myr) & ($10^{3}$K) & (km s$^{-1}$) & ($\MsunPerYear$) & (pc) & \ldots{} & \ldots{} & \tabularnewline
\hline 
$3.0^{a}$ & 2.8$^{a}$ & 911.2$^{b}$ & B9 IV$^{e,f}$ & $0.4^{b}$ & 10.5$^{b}$  & 118$^{f}$ & $3.5\times10^{-7}\,^{a}$ & $318_{-46}^{+65}\,^{i}$ & $5.4^{k}$ & $6.1^{k}$ & $0.4^{e}$\tabularnewline
$4.2^{c}$ &  & 302$^{c}$ &  & $0.16^{c}$ & 10.7$^{c}$ & 280$^{g}$ & $1.4\times10^{-5\, h}$ & $280_{-50}^{+80}\,^{c,j}$ &  &  & $0.5^{c}$\tabularnewline
 &  & $295^{d}$ &  &  &  &  &  & $543^{b}$ &  &  & \tabularnewline
\hline 
\end{tabular}

\end{center}

$^{a}$\citet{Brittain:2007}, $^{b}$\citet{Montesinos:2009}, $^{c}$\citet{VanDenAncker:1998},
$^{d}$\citet{Monnier:2005}, $^{e}$\citet{Malfait:1998}, $^{f}$\citet{Mora:2001},
$^{g}$\citet{Grady:1996}, $^{h}$\citet{Mendigutia:2011}, $^{i}$\citet{VanLeeuwen:2007},
$^{j}$\citet{Perryman:1997}, $^{k}$\citet{Cutri:2003}.

\end{table*}
%%%%%%%%%%%%%%%%%%%%%%%%%%%%%%%%%%%%%%%%%%%%%%%%%%%%%%%%%

Strong outflows are commonly associated with the early stages of
stellar evolution. They are likely responsible for transporting excess
angular momentum away from the star-disc system and regulating the
mass-accretion process and spin evolution of newly born stars
(e.g.~\citealt{hartmann:1989}; \citealt{matt:2005};
\citealt{Bouvier:2014}). There are at least three possible types of
outflows around young stellar objects: (1)~a disc wind launched from
the accretion disc (e.g.~\citealt{blandford:1982};
\citealt{ustyugova:1995}; \citealt{Romanova:1997};
\citealt{ouyed:1997}; \citealt{Ustyugova:1999};
\citealt{koenigl:2000}; \citealt{Pudritz:2007}), (2)~an X-wind or a
conical wind launched near  
the disc-magnetosphere interaction region
(e.g.~\citealt{shu:1994,Shu:1995}; \citealt{Romanova:2009}) and (3) a
stellar wind launched from open magnetic field lines anchored to the
stellar surface (e.g.~\citealt{decampli:1981};
\citealt*{hartmann:1982}; \citealt{Kwan:1988}; \citealt{Hirose:1997};
\citealt{Strafella:1998}; \citealt{Romanova:2005};
\citealt{matt:2005}; \citealt{Cranmer:2009}).  However, despite recent
efforts, the exact launching mechanisms of the winds and outflows as
well as the mechanism behind the collimation of the ejected gas into
jets are still not well understood (e.g.~\citealt{Edwards:2006};
\citealt*{Ferreira:2006}; \citealt*{Kwan:2007};
\citealt{Tatulli:2007b}; \citealt{Kraus:2008}; 
\citealt*{Eisner:2014}). Hence, high-resolution interferometric
observations which can resolve the wind launching regions are crucial
for addressing this issue. If a high spectral resolution is combined
with a high spatial resolution, the emission-line regions near the
base of the wind (e.g., Br$\gamma$ emission regions in Herbig Ae/Be
stars) can be resolved in many spectral channels across the line
(e.g.~\citealt{Weigelt:2011,Kraus:2012,Kraus:2012b,GarciaLopez:2015}). This allows
us to study the wavelength-dependent extent and the kinematics of the
winds, which can be derived from the line visibilities and
wavelength-dependent differential and closure phases. Such
observations would help us to distinguish the different types of
outflow scenarios.

HD~58647 is a bright (K=5.4, H=6.1) Herbig Be star (B9 IV:
\citealt*{The:1994}; \citealt*{Malfait:1998}; \citealt{Mora:2001}) which
is located at 318\,pc (\citealt{VanLeeuwen:2007}; hereafter VL07).
The star exhibits double-peaked profiles in some hydrogen emission
lines, such as H$\alpha$ and Br$\gamma$ (e.g.~\citealt{Grady:1996};
\citealt{Oudmaijer:1999}; \citealt{Harrington:2009};
\citealt{Brittain:2007}). The star does not show a significant
variability in H$\alpha$ on time scales of 3\,d
(\citealt{Mendigutia:2011}) and 1\,yr (\citealt{Harrington:2009}). A
presence of an intrinsic linear polarization in H$\alpha$ is reported
in \citet{Vink:2002}, \citet{Mottram:2007} and
\citet{Harrington:2009}.  A relatively strong double-peaked Br$\gamma$
is reported in \citet{Brittain:2007}; however, no variability
study is found for this line in the literature.  The age of the star has been
estimated as 0.4\,Myr (\citealt*{VanDenAncker:1998}) and 0.16~Myr
(\citealt{Montesinos:2009}), but some have also reported the
possibilities of the star being an older classical Be star
(e.g.~\citealt*{Manoj:2002}; \citealt{Berthoud:2007}).
Using the broad K-band (2.18\,$\micron$, $\Delta \lambda=0.3\,\micron$) Keck
interferometer and a uniform ring geometric model, \citet{Monnier:2005}
(hereafter MO05)
estimated the ring radius of the K-band continuum emission of HD~58647 as
0.82($\pm0.13$)\,au, assuming the distance to the star is 280~pc
(\citealt{VanDenAncker:1998}; \citealt{Perryman:1997}). However, this
radius may be underestimated since the ring inclination effect is not
included in their study.  Some of the known properties of HD~58647 are
summarised in Table~\ref{tab:published-parameters}.

Here, we present VLTI-AMBER observations of the Herbig Be star
HD~58647 with the high spectral resolution mode
(HR-K-2.172) to resolve its Br$\gamma$ emission spatially and
spectrally ($\mathrm{R}\approx12000$, $\Delta v\approx25\,\mathrm{km\,
  s^{-1}}$). The velocity extent of the Br$\gamma$ line profile for
the star is about $\pm200\,\mathrm{km\, s^{-1}}$
(\citealt{Brittain:2007}); hence, this 
resolution will provide us with more than 16 velocity components
across the line, which is essential for studying wind kinematics in a
subsequent radiative transfer modelling.  
So far, there are only a few Herbig stars which are observed with the
VLTI-AMBER in the high spectral resolution mode: 
Z~CMa (Be + FUor, \citealt{Benisty:2010}), 
MWC~297 (B1.5\,V, \citealt{Weigelt:2011}), 
HD~163296 (A1\,V, \citealt{GarciaLopez:2015}), HD~98922 (B9\,V,
\citealt{Caratti:2015}) and HD~100564 (B9\,V,
\citealt{Mendigutia:2015}). \citet{Kraus:2012b} and 
\citet{Ellerbroek:2015} have presented high spectral resolution
VLTI-AMBER observations of V921 Sco (B[e]) and HD~50138 (B[e]),
respectively, but their evolutionary states are not well known. 
These studies have  successfully demonstrated the
feasibility and usefulness of this type 
of observation for probing the kinematics and origin of the wind from
Herbig stars. One notable difference between HD~58647 and other
objects (MWC~297, HD~163296 and HD~98922) is the shape of  Br$\gamma$. The
former has a double-peaked and the latter have a single-peaked shaped
line profile.  It is interesting to study whether the
difference is caused simply by the inclination angle effect or due to
a fundamental difference in their wind structures. For example, the
Br$\gamma$ emission could originate from a bipolar/stellar wind or 
a disc wind (e.g.~\citealt{Malbet:2007}; \citealt{Tatulli:2007b};
\citealt{Kraus:2008}; \citealt{Eisner:2010}; \citealt{Weigelt:2011}).
If the line is double-peaked, the emission is most likely  from 
a disc wind, which is rotating near Keplerian velocity at the base of
the wind. If the line is single-peaked, it is likely formed in either 
a bipolar/stellar wind viewed at a mid to high inclination angle or a disc
wind viewed at a low inclination angle (e.g.\,\citealt{Kurosawa:2006}). 

The interferometric data will be analysed by using the radiative
transfer code \textsc{torus} (e.g., 
\citealt{Harries:2000}; \citealt*{Symington:2005};
  \citealt*{Kurosawa:2006}; \citealt*{Kurosawa:2011}, hereafter KU11;
  \citealt{Haworth:2012}), which calculates 
hydrogen emission line profiles and line intensity maps using various
inflow/outflow models of young stars. The code has been used to study
classical T~Tauri stars (\citealt*{Kurosawa:2005};
\citealt*{Kurosawa:2008};
\citealt{Kurosawa:2012,Alencar:2012,Kurosawa:2013}), a pre-FUor
(\citealt{Petrov:2014}) and a Herbig Ae star
(\citealt{GarciaLopez:2016}) in the past.  Based on the simulated
emission maps computed in many velocity bins, we can reconstruct not
only the line profiles, but also important interferometric quantities
such as visibility, differential and closure phases across the line,
which can be directly compared with the observations. To differentiate
various possible outflow scenarios (e.g., a stellar wind and a disc
wind), we will perform qualitative analysis of wind structures around
YSOs by detailed radiative transfer models for the high resolution
spectro-interferometric observations.

The aim of this paper is to investigate the nature and the origin of
the wind around the Herbig Be star HD~58647 by means of high
spectral resolution interferometric observations combined with
radiative transfer modellings.  In Section~\ref{sec:observations}, we
describe our VLTI-AMBER observations and data reduction. A simple
geometrical model to characterise the size of continuum emission 
is presented in Section~\ref{sec:ring-model-fit}. The description of
the radiative transfer models and the results of model fits to the
observations are given in Section~\ref{sec:RT-models}. Brief
discussions on our results are given in
Section~\ref{sec:discussion}. Finally, the summary of our findings and
conclusions are presented in Section~\ref{sec:conclusions}.

%%%%%%%%%%%%%%%%%%%%%%%%%%%%%%%%%%%%%%%%%%%%%%%%%%%%%%%%%
\begin{figure}
\begin{center}

\includegraphics[clip,width=0.40\textwidth]{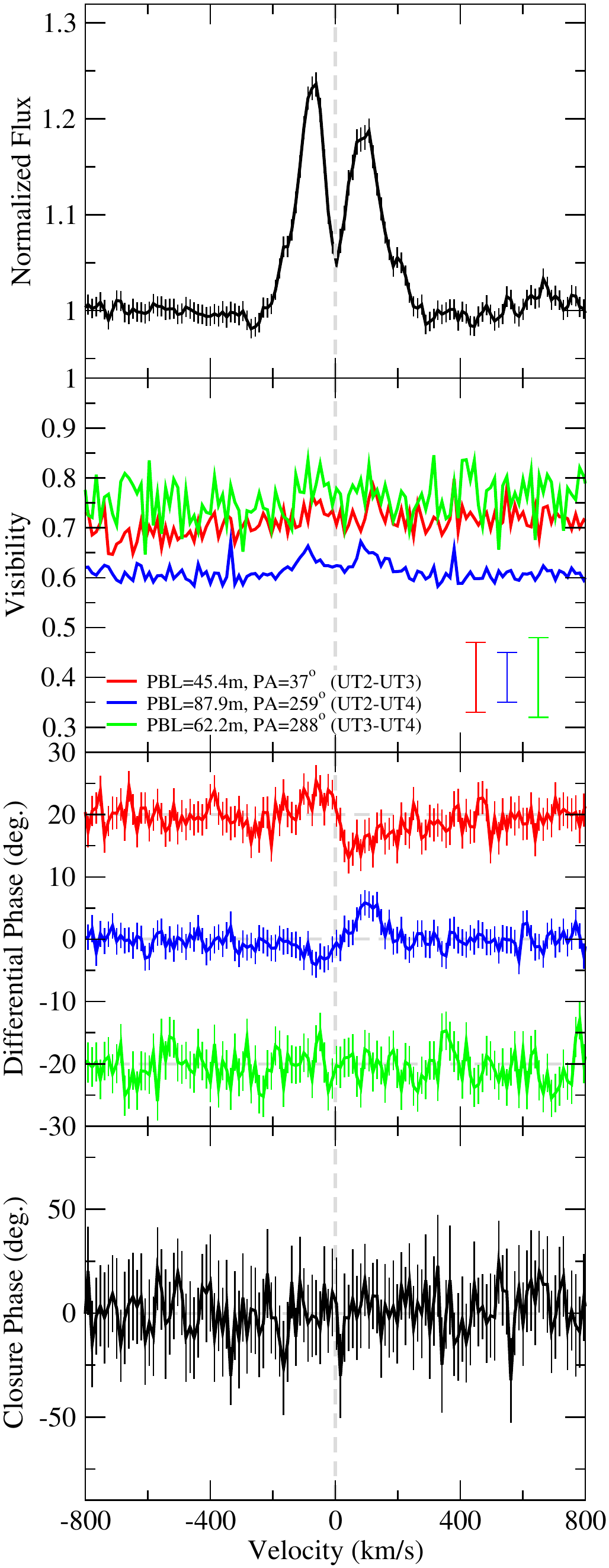}

\caption{Summary of the high-spectral-resolution AMBER observations around Br$\gamma$ using three
  baselines. The figure shows the average normalised flux around
  Br$\gamma$ from 3 telescopes (top panel), the visibilities (second panel
  from the top), differential phases (third panel from the top) and closure
  phases (bottom panel) as a function of velocity. For clarity, the
  differential phases for the first (red) and third (green) 
  baselines are shifted vertically by $+20^{\circ}$ and $-20^{\circ}$,
  respectively. Thin vertical lines indicate the 
  uncertainty in each data point. In the visibility plot (the second
  panel from the top), the typical uncertainty levels are indicated
  on the lower-right corner. \label{fig:amber-summary}}

\end{center}

\end{figure}
%%%%%%%%%%%%%%%%%%%%%%%%%%%%%%%%%%%%%%%%%%%%%%%%%%%%%%%%%

%%%%%%%%%%%%%%%%%%%%%%%%%%%%%%%%%%%%%%%%%%%%%%%%%%%%%%%%%
\begin{table*}

\caption{Log for high-spectral-resolution VLTI-AMBER observation of
  HD\,58647 using the telescope configuration UT2-UT3-UT4. \label{tab:obs-log-high-res}}

\begin{center}

\begin{tabular}{ccccccccc}
\hline 
Object & Type & Date & Time {[}UT{]}  & Mode & $\lambda$ Range & DIT$^{a}$ & NDIT$^{b}$ & Seeing\tabularnewline
 &  &  &   &  & ($\micron$) & (s) &  & ('')\tabularnewline
\hline 
HD~58647 & Target & 2015-01-02 & 04:31 -- 05:13  & HR-K-F & 2.147--2.194 & 0.75 & 2400 & 0.75 -- 1.0\tabularnewline
HD~57939 & Calibrator & 2015-01-02 & 05:24 -- 05:35  & HR-K-F & 2.147--2.194 & 0.75 & 600 & 0.87 -- 1.1\tabularnewline
\hline 
\end{tabular}

$^{a}$Detector integration time per interferogram. $^{b}$Total number
of interferograms.

\end{center}

\end{table*}
%%%%%%%%%%%%%%%%%%%%%%%%%%%%%%%%%%%%%%%%%%%%%%%%%%%%%%%%%

%%%%%%%%%%%%%%%%%%%%%%%%%%%%%%%%%%%%%%%%%%%%%%%%%%%%%%%%% 
\begin{table*}

\caption{Log for low-spectral-resolution VILT-AMBER observations of
  HD\,58647. \label{tab:obs-log-low-res}}

\begin{center}
\begin{tabular}{ccccccccc}
\hline 
Data set & UT date & UT time & DIT & NDIT & Telescopes &
Proj.~baselines & PA & ESO Prog.\,ID\tabularnewline
 & (yyyy:mm:dd)  & (hh:mm) & (s) &  & (ATs) & (m) & ($^{\circ}$) &\tabularnewline
\hline 
A & 2009-11-11 & 08:24 & 0.05 & 5000 & D0-K0-H0 & 63.5/31.7/95.2 & 70.2/70.1/70.2 & 084.C-0170\tabularnewline
B & 2009-11-11 & 09:07 & 0.05 & 2000 & D0-K0-H0 & 64.0/32.0/95.9 & 72.8/72.8/72.8 & 084.C-0170\tabularnewline
C & 2009-11-12 & 08:00 & 0.05 & 2500 & D0-G1-H0 & 71.5/62.5/70.2 & 131.2/68.5/3.5 & 084.C-0170\tabularnewline
D & 2009-11-12 & 08:22 & 0.05 & 1500 & D0-G1-H0 & 71.3/63.5/70.3 & 131.7/70.2/5.9 & 084.C-0170\tabularnewline
E & 2010-12-31 & 03:07 & 0.05 & 5000 & I1-H0-G0 & 26.6/40.1/50.5 & 57.5/141.0/109.5 & 086.C-0267\tabularnewline
F & 2010-12-31 & 03:41 & 0.05 & 5000 & I1-H0-G0 & 28.6/40.6/53.5 & 62.0/142.2/110.4 & 086.C-0267\tabularnewline
G & 2011-02-12 & 01:59 & 0.10 & 5000 & I1-H0-G1 & 40.7/45.3/70.2 & 146.2/37.0/3.8 & 086.C-0267\tabularnewline
\hline 
\end{tabular}

\end{center}

\end{table*}
%%%%%%%%%%%%%%%%%%%%%%%%%%%%%%%%%%%%%%%%%%%%%%%%%%%%%%%%% 

\section{Observations and data reductions}

\label{sec:observations}

The Herbig Be star HD58647 (B9 IV) was observed during the second half
of the night of January 1, 2015 using the Very Large Telescope
Interferometer (VLTI) with the AMBER beam combiner instrument
(\citealt{Petrov:2007}), as a part of the observing programme
094.C-0164 (PI: Kurosawa). The combination of UT2-UT3-UT4 telescopes
along with the high spectral resolution mode (R=12000) of AMBER were
used. The log of the observations is summarised in
Table~\ref{tab:obs-log-high-res}. The interferograms were obtained
using the fringe tracker FINITO for co-phasing with the detector
integration time (DIT) of 0.75~s for each interferogram.  The total of
2400 interferograms were recorded, or equivalently about 30 min of
total integration time was used. In addition to our target, HD~57939
(K0~III) was observed with the same DIT, as an interferometric
calibrator.

The fringe tracking performance typically varies between the target
and calibrator observations; therefore, low-resolution (LR, R
$\approx$ 35) AMBER observations (ESO Prog. ID: 084.C-0170 and
086.C-0267, PI: Natta) are used to calibrate the continuum
visibilities. A summary of the LR observations is shown in 
Table~\ref{tab:obs-log-low-res}.  The data were reduced with
the AMBER data processing software \textsc{amdlib} (v3.0.5)
(\citealt{Tatulli:2007a}; \citealt*{Chelli:2009}), which is available
at \url{http://www.jmmc.fr/data_processing_amber.htm}.
Furthermore, for the reduction of the LR data, we applied a  
software that equalizes the optical path difference
(OPD) histograms of the object and calibrator (\citealt{Kreplin:2012}).
The end products of the data reduction
process are the wavelength-dependent normalised flux (averaged over 3
telescopes), visibilities, differential phases, and closure
phases. The typical errors in the observed visibilities are
approximately 5 to 9~per~cent. The results are summarised in
Fig.~\ref{fig:amber-summary}.

%%%%%%%%%%%%%%%%%%%%%%%%%%%%%%%%%%%%%%%%%%%%%%%%%%%%%%%%%
\begin{figure}
\begin{center}
\includegraphics[clip,width=0.46\textwidth]{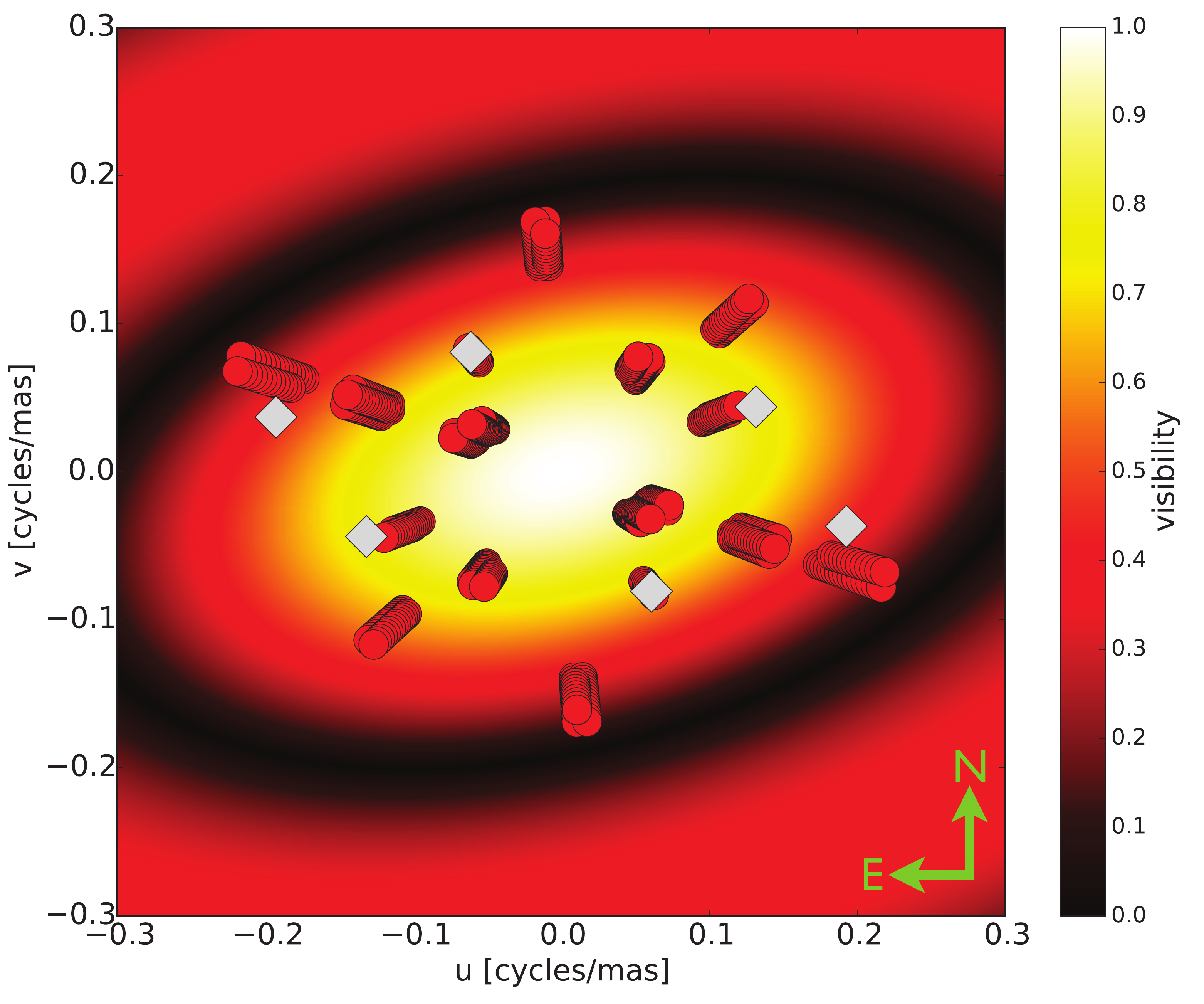}
\end{center}

\caption{The $uv$ points used for  
  the low-spectral-resolution (filled circle,
  Tabel~\ref{tab:obs-log-low-res}) and high-spectral-resolution
  (filled diamond, Tabel~\ref{tab:obs-log-high-res}) AMBER
  observations are overplotted with the visibility from a geometrical
  ring model.  The parameters used for the ring model are
  summarised in Table~\ref{tab:ring-fit-param}.  The position angles of the
  baselines are measured from north to
  east. Here, \,$u$ decreases towards east which is the opposite 
  to the sense used in the usual convention (e.g.~\citealt*{Thompson:2001}). 
  \label{fig:uv-coverage}} 
\end{figure}
%%%%%%%%%%%%%%%%%%%%%%%%%%%%%%%%%%%%%%%%%%%%%%%%%%%%%%%%%

%%%%%%%%%%%%%%%%%%%%%%%%%%%%%%%%%%%%%%%%%%%%%%%%%%%%%%%%% 
\begin{figure}

\begin{center}
\includegraphics[clip,width=0.44\textwidth]{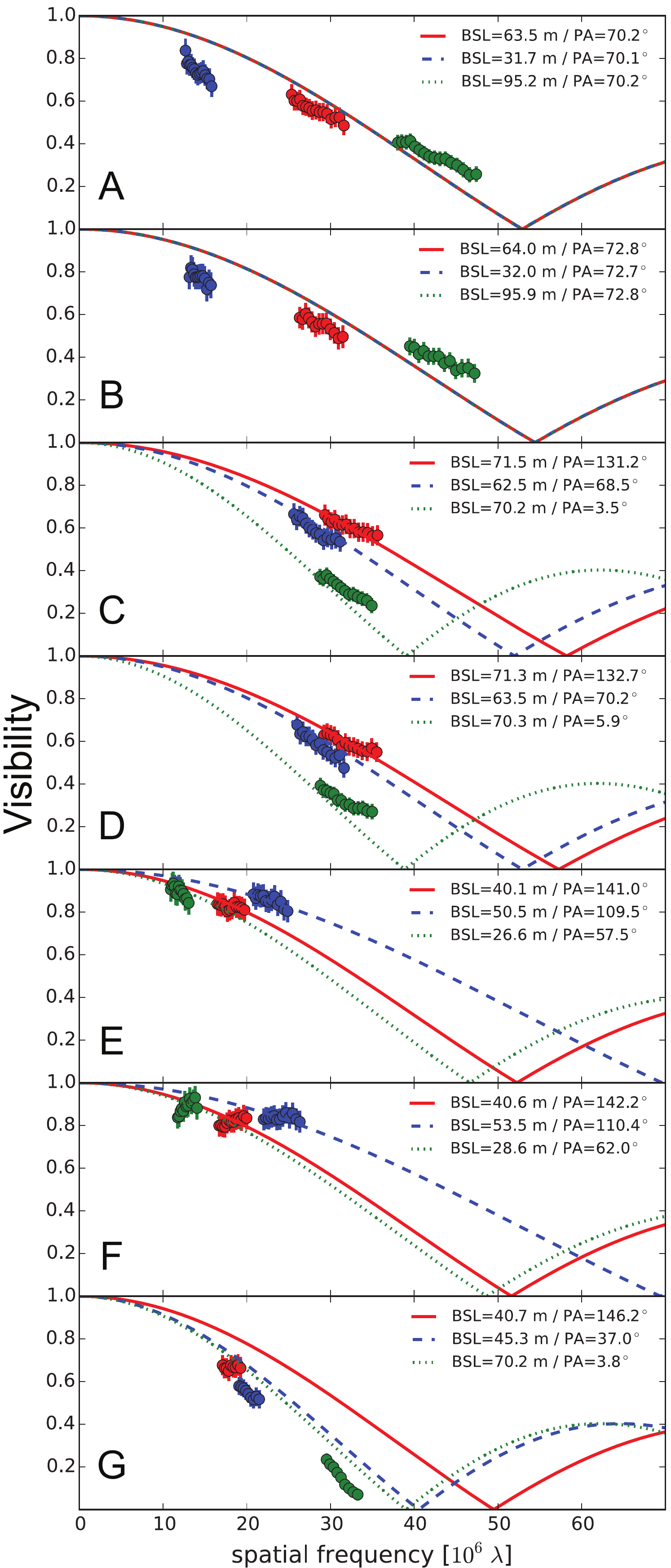}
\end{center}

\caption{Summary of a geometrical ring model fit for the circumstellar
  visibilities ($V_\mathrm{circ}$) from the low-resolution (LR,
  $R\approx 35$) VLTI-AMBER observations
  (Table~\ref{tab:obs-log-low-res}). The circumstellar visibilities
  from 7 different sets of measurements (Sets A--G in
  Table~\ref{tab:obs-log-low-res} from top to bottom) are shown as a
  function of spatial frequency (small circles with error bars). The
  dashed, dotted and solid lines are the model visibility curves of
  the two-dimensional (elongated ring) model at the same PAs as in
  the observations. The corresponding parameters used in the best-fit
  model are summarised in Table~\ref{tab:ring-fit-param}. The
  corresponding $uv$ coverage of the LR VLTI-AMBER observations are
  shown in Fig.~\ref{fig:uv-coverage}.  Note that all the model
  visibility curves nearly overlap in the top two panels, thus they
  are hard to distinguish from the others.  \label{fig:ring-fit}}

\end{figure}
%%%%%%%%%%%%%%%%%%%%%%%%%%%%%%%%%%%%%%%%%%%%%%%%%%%%%%%%%

The figure clearly shows that Br$\gamma$ is double peaked with its
peaks at around 25~per~cent above the continuum. The dip is located at
the line centre. The double-peaked line profiles are likely caused by
the rotation of the gas around the star, most likely in a disc wind
(e.g.~\citealt{Shlosman:1993,
  knigge:1995,sim:2005b,Kurosawa:2006}). This will be further
investigated in Section~\ref{sec:RT-models}. The wavelength-dependent
visibilities are rather noisy, but those from the longest projected
baseline (87.9~m) clearly show that the visibility level increases
in the line. The visibility curve also shows a double-peak appearance. The
noise levels in the visibility curves from the shorter 
baselines (45.4 and 62.2~m) are somewhat higher; therefore, the
double-peak feature is only weakly suggested. The increases of the
visibilities in the emission line with respect to those in the
continuum indicate that the size of line-emitting region is smaller
than that of the K-band continuum-emitting region.

The wavelength-dependent differential phases show S-shape curves
around the line centre for two projected baselines (45.4 and 87.9~m).
In both cases, the differential phase crosses 0 near the line
centre.  The sense of the S-shape pattern is opposite between the
two. The deviation of the differential phases from $0^{\circ}$
indicates a shift in the photocentre at the velocity channel. Since
the pattern of the differential phase curves is sensitive to the
kinematics and geometry of the line-emitting gas, this will be used to
constrain the gas outflow models used in the radiative transfer
calculations which are to be presented later in
Section~\ref{sub:model-vis}. Lastly, the wavelength-dependent closure
phases do not show any significant deviation from $0^{\circ}$ within
the range of uncertainties which are typically around $\pm20^{\circ}$.

%%%%%%%%%%%%%%%%%%%%%%%%%%%%%%%%%%%%%%%%%%%%%%%%%%%%%%%%% 
\begin{table}

\caption{Parameters of the best-fit geometrical ring model for
  the low-resolution AMBER visibilities. 
  \label{tab:ring-fit-param}}

\begin{center}

\begin{tabular}{cccccc}
\hline 
Model & $R_{\mathrm{ring}}$ &$i$ & PA \tabularnewline
      & (mas) & ($^{\circ}$) & ($^{\circ}$)\tabularnewline
\hline 
Ring  & $2.0(\pm 0.3)$ &$55(\pm 2)$ & $105(\pm 4)$\tabularnewline
\hline 
\end{tabular}
\end{center}

\end{table}
%%%%%%%%%%%%%%%%%%%%%%%%%%%%%%%%%%%%%%%%%%%%%%%%%%%%%%%%% 

\section{A simple geometrical model for the continuum emission region}

\label{sec:ring-model-fit}

To estimate the characteristic size of the K-band continuum emission
of HD~58647, we fit the observed visibilities of the LR VLTI-AMBER
observations, listed in Table~\ref{tab:obs-log-low-res}, with a simple
geometrical ring model. Since the observed visibilities
($V_{\mathrm{obs}}$) are influenced by both unresolved stellar flux
($F_{*}$) and that from the circumstellar matter
($F_{\mathrm{circ}}$), we first calculate the circumstellar
visibilities ($V_{\mathrm{circ}}$) as follows:
\begin{equation}
V_{\mathrm{circ}}=\frac{\left|V_{\mathrm{obs}}\left(F_{*}+F_{\mathrm{circ}}\right)-F_{*}V_{*}\right|}{F_{\mathrm{circ}}}\label{eq:vis-circumstellar}
\end{equation}
where $V_{*}$ is the stellar visibility (e.g.~\citealt{Kreplin:2012}).
Here, we assume $V_{*}=1$ since the central star is unresolved (its
angular radius is $\sim$0.07~mas at the distance of 318~pc, 
VL07). To estimate the ratio of the circumstellar
to stellar fluxes, we compare the stellar atmosphere model of \citet{Kurucz:1979}
with $T_{\mathrm{eff}}=10500\,$K and $\log g_{*}=3.5$ (\citealt{Montesinos:2009}) with the K-band
photometry data from VO SED Analyzer tool (VOSA, \citealt{Bayo:2008}),
after dereddening the data ($A_{V}=0.4$ , \citealt{Malfait:1998}). 
The original photometric data are from
\citet*{Ochsenbein:2000}. The target distance ($d\sim318$~pc, VL07) 
is adopted from the \textit{Hipparcos} measurement. We find the flux
ratio is $F_{\mathrm{circ}}/F_{*}=2.9$ 
at $\mbox{\ensuremath{\lambda}=}2.2\,\micron$. Using this value and
equation~\ref{eq:vis-circumstellar}, the circumstellar visibilities
$V_{\mathrm{circ}}$ are determined for each set of the LR VLTI-AMBER
observations (Table~\ref{tab:obs-log-low-res}). The results are shown in
Fig.~\ref{fig:ring-fit}.  The figure suggests that the emission
geometry deviates from a spherical symmetry because e.g.~the
circumstellar visibilities at the spatial frequency $\sim30\times 10^{6}\,\lambda$
(e.g. in the data sets~C and D) are notably different for the
baselines with different position angles. 

To model the circumstellar visibilities $V_{\mathrm{circ}}$, we
consider an inclined uniform ring geometry with an inner radius
$R_{\mathrm{ring}}$ and a width $\Delta R_{\mathrm{ring}}$ because the
K-band continuum of the discs of Herbig stars is expected to originate
mainly from the region near the dust sublimation radius (e.g.~MO05;
\citealt{Dullemond:2010}).  Here, we assume $\Delta
R_{\mathrm{ring}}=0.2\, R_{\mathrm{ring}}$ as in MO05. The elongation
of the ring occurs when the circular ring has an inclination angle
$i>0^{\circ}$.  We use this two-dimensional model to fit the set of
all visibilities from the LR observations simultaneously, instead
of fitting one-dimensional model visibilities to individual observed
visibilities.  The results of the best-fit model are shown in
Fig.~\ref{fig:ring-fit}. The corresponding model parameters are
summarised in Table~\ref{tab:ring-fit-param}. As one can see from the
figure, the simple geometrical model with a ring can reasonably fit
the observed visibilities. In this analysis, we find that the angular
radius of the K-band continuum-emitting ring is around 2.0~mas
(Table~\ref{tab:ring-fit-param}), which corresponds to 0.64~au at the
distance of 318~pc (VL07).

\section{Radiative Transfer Models}

\label{sec:RT-models}

To model the interferometric quantities obtained by VLTI-AMBER in
the high-spectral-resolution mode, i.e.~the wavelength-dependent flux,
visibility and differential and closure phases around Br$\gamma$
presented in Section~\ref{sec:observations}, we use the radiative
transfer code {\sc torus} (e.g.~\citealt{Harries:2000, Symington:2005,
  Kurosawa:2006}; KU11).  The numerical method used in the
current work is
essentially the same as in KU11. The model uses the
tree-structured grid (octree/quadtree in 3-D and 2-D, respectively) in
cartesian coordinate, but here we assume axisymmetry around the
stellar rotation axis.  The atomic model consists of hydrogen with 20
energy levels, and the non-local thermodynamic equilibrium (non-LTE)
level populations are solved using the Sobolev approximation
\citep[e.g.][]{Sobolev:1957,castor:1970,castor:1979}. More
comprehensive descriptions of the code are given in KU11.

For the present work, we have made three minor modifications to the code
in KU11: (1)~the model now includes the continuum
emission from a geometrically flat ring which simulates the K-band dust
emission near the dust sublimation radius, (2)~the effect of rotation
of a magnetosphere has been included by using the method described in
\cite*{Muzerolle:2001}, and (3)~the code has been modified to write
model intensity maps as a function of wavelength (or velocity
bins). The first modification is important for modelling a correct
line strength of Br$\gamma$ (2.2166~$\mu$m) and the interferometric
quantities around the line. The second modification is included
because intermediate-mass Herbig Ae/Be stars are expected to be
rotating relatively fast with their rotation periods about $0.2$ to a few
days (e.g.~\citealt{Hubrig:2009}, but see also
\citealt{Hubrig:2011}). Until now,
our model has been applied mainly to classical T Tauri stars which
rotate relatively slowly with their rotation period 2--10\,d
(e.g.~\citealt{Herbst:1987}; \citealt{Herbst:1994}) in which  the effect
of rotation is rather small (e.g.~\citealt{Muzerolle:2001}). The
third modification is important because it will allow us to compute
the interferometric quantities, 
such as visibility, differential and closure phases, as a function of
wavelength. This enables us to directly compare our models with the
spectro-interferometric data from VLTI-AMBER.

\subsection{Model configurations }
\label{sub:model-config}

\begin{figure}
\begin{center}

\includegraphics[clip,width=0.4\textwidth]{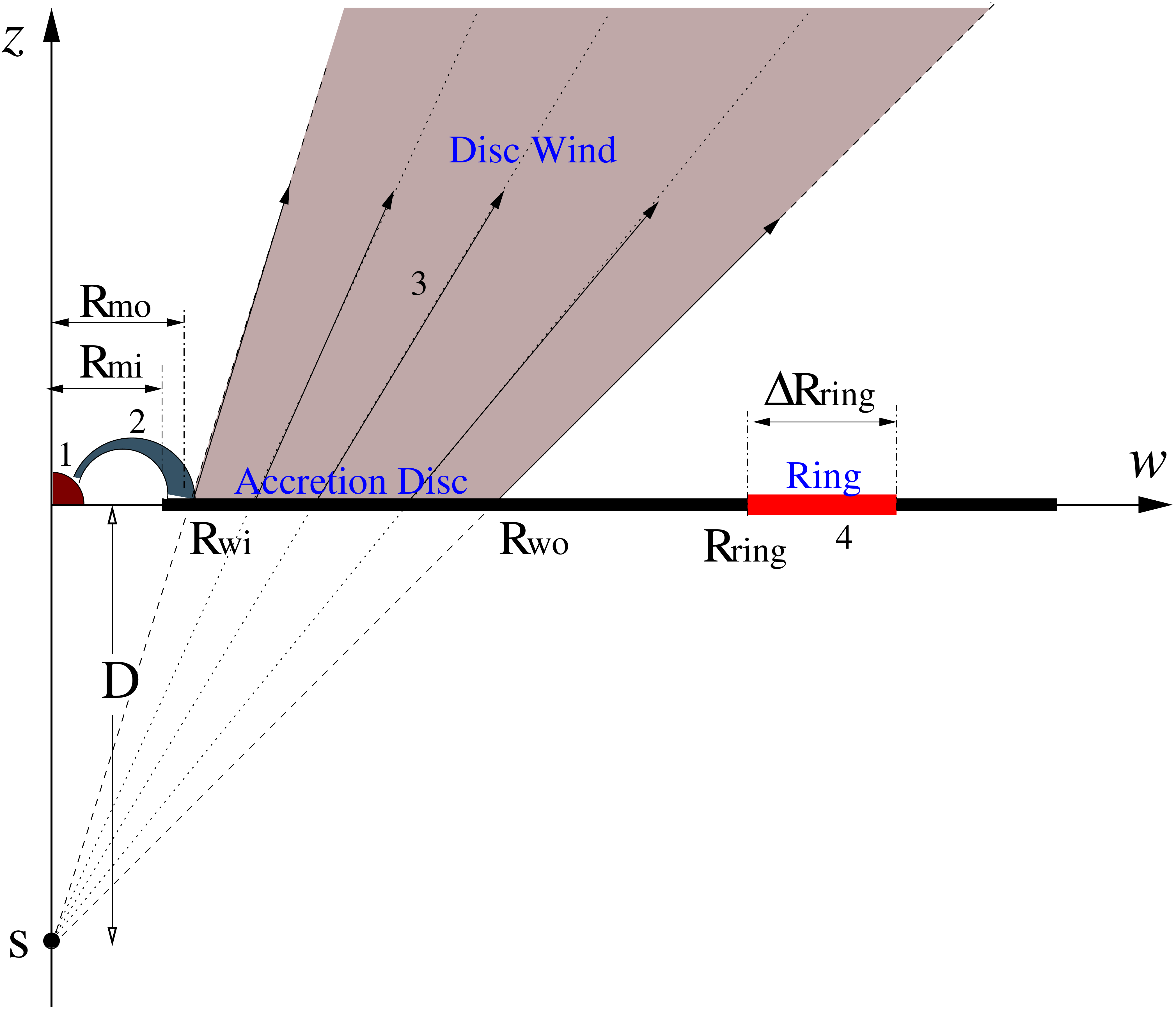}

\caption{The model configuration of the circumstellar materials around
  the Herbig Be star HD~58647.  The density is assumed
  to be axisymmetric and top-down symmetric. The model consists
  of: (1)~the stellar continuum source (star) located at the origin of 
  the cartesian coordinates system ($w$,$z$), (2)~the magnetospheric accretion
  flow, (3)~the disc wind, (4)~the K-band continuum-emitting ring, 
  and (5)~the geometrically thin accretion disc. The split-monopole (with the source
  displacement distance $D$) disc wind emerges from the equatorial
  plane, but only within the distances from $z$-axis between
  $R_{\mathrm{wi}}$ and $R_{\mathrm{wo}}$. The geometrically thin ring
  with its inner radius $R_{\mathrm{ring}}$ and width $\Delta
  R_{\mathrm{ring}}$ is also placed on
  the equatorial plane. See Section~\ref{sub:model-config} for more
  information.  The figure is not to scale. \label{fig:model-config} }
\end{center}

\end{figure}

There are 4 distinct emission components in our model: (1)~the central
star, (2)~the magnetospheric accretion funnels as described by
\citet*{Hartmann:1994} and \citet{Muzerolle:2001}, (3)~the disc wind
which emerges from the equatorial plane (a geometrically thin
accretion disc) located outside of the magnetosphere and (4)~the
geometrically flat ring which simulates the K-band dust emission near
the dust sublimation radius.  A basic schematic diagram of our models
is shown in Fig.~\ref{fig:model-config}.  In the following, we
briefly describe each component and the key parameters. More
comprehensive descriptions of the magnetosphere and disc wind
components can be found in KU11.

%************************************************************
%************************************************************

\subsubsection{Central star}

\label{subsub:Central-star}

The central star is the main source of continuum radiation which ionizes the
gas in the flow components (a magnetospheric accretion and a disc
wind).  For a given effective temperature $T_{\mathrm{eff}}$ and its
surface gravity $\log g_*$, we use the stellar atmosphere model of
\citet{Kurucz:1979} as the photospheric contribution to the 
continuum in our model.  Our underlying central star is a late B-type star;
hence, the photospheric absorption component from the star may affect
the shape of the model Br$\gamma$ profiles. For this reason, we use
the high-spectral-resolution synthetic \textsc{phoenix}
(\citealt{Hauschildt:1999}) spectra from \citet{Husser:2013},
convolved with $v\sin i=118\,\kmps$ (\citealt{Mora:2001}) for the
photospheric absorption profile for Br$\gamma$. 

\subsubsection{Magnetosphere }

\label{subsub:MA-config}

We adopt a dipole magnetic field geometry with its accretion
stream described by $r=R_{\mathrm{m}}\,\sin^{2}{\theta}$ where $r$ and
$\theta$ are the polar coordinates; $R_{\mathrm{m}}$ is the
magnetospheric radius on the equatorial plane, as in
e.g.~\cite{ghosh:1977} and \cite{Hartmann:1994}. In this model, the
accretion occurs in the funnel regions defined by two stream lines
which correspond to the inner and outer magnetospheric radii, i.e.,
$R_{\mathrm{m}}=R_{\mathrm{mi}}$ and $R_{\mathrm{mo}}$
(Fig.~\ref{fig:model-config}).  As in KU11, we follow
the density and temperature structures along the stream lines as in
~\citet{Hartmann:1994} and \cite{Muzerolle:2001}. The temperature
scale is normalised with a parameter $T_{\mathrm{m}}$ which sets the
maximum temperature in the stream. The mass-accretion rate
$\dot{M}_{\mathrm{a}}$ scales the density of the magnetospheric
accretion funnels. We adopt the method described in 
\cite{Muzerolle:2001} (their Appendix A), to include the effect of rotation in the
accretion funnels.

\subsubsection{Disc wind}

\label{subsub:DW-config}

As in KU11, the disc wind model used here is based on
the `split-monopole' wind model by e.g.~\citet{knigge:1995} and
\citet{Long:2002} (see also \citealt{Shlosman:1993}). The outflow
arises from the surface of the rotating accretion disc, and has a
bi-conical geometry.  Most influential parameters in our models are:
(1)~the total mass-loss rate in the disc wind ($\dot{M}_{\mathrm{dw}}$),
(2)~the degree of the wind collimation, (3)~the steepness of the
radial velocity ($\beta_{\mathrm{dw}}$ in
equation~\ref{eq:discwind-poloidal-velocity} below), and 
(4)~the wind temperature. Fig.~\ref{fig:model-config} shows the basic
configuration of the disc-wind model. The wind arises from the disc
surface. The wind ``source'' points ($S$), from which the stream lines
diverge, are placed at distance $D$ above and below the centre of the
star. The angle of the wind launching is controlled by changing the
value of $D$. The wind launching is confined between $R_{\mathrm{wi}}$
and $R_{\mathrm{wo}}$ where the former is normally set to the outer
radius of the closed magnetosphere ($R_{\mathrm{mo}}$) and the latter
to be a free parameter.  The wind density is normalised with the total
mass-loss rate in the disc wind ($\dot{M}_{\mathrm{dw}}$) while the
local mass-loss rate per unit area ($\dot{m}$) follows the power-law
$\dot{m} \propto w^{-p}$ where $w$ is the distance from the star on the
equatorial plane and $p=2.4$ (our fiducial value which produces
reasonable Br$\gamma$ line profiles). 

The poloidal component of the wind velocity ($v_{\mathrm{p}}$) is parametrised as: 
\begin{equation}
  v_{\mathrm{p}}\left(w_{i},l\right)=c_{\mathrm{s}}\left(w_{i}\right)+\left[f\,
    v_{\mathrm{esc}}-c_{\mathrm{s}}\left(w_{i}\right)\right]
    \left(1-\frac{R_{\mathrm{s}}}{l+R_{\mathrm{s}}}\right)^{\beta_{\mathrm{dw}}}
  \label{eq:discwind-poloidal-velocity}
\end{equation} 
where $w_{i}$, $c_{\mathrm{s}}$, $f$, and $l$ are the distance from
the rotational axis ($z$) to
the wind emerging point on the disc, the sound speed at the
wind launching point on the disc, the constant scale factor of the
asymptotic terminal velocity to the local escape velocity ($v_{\mathrm{esc}}$
from the wind emerging point on the disc), and the distance from the
disc surface along stream lines respectively. $R_{s}$ is the wind
scale length, and is set to 10~$R_{\mathrm{wi}}$ by following
\citet{Long:2002}. 
The azimuthal component of the wind velocity $v_{\phi}\left(w,z\right)$
is computed from the Keplerian rotational velocity at the emerging
point of the stream line i.e. $v_{\phi}\left(w_{i},0\right)=\left(GM_{*}/w_{i}\right)^{1/2}$
by assuming the conservation of
the specific angular momentum along a stream line, i.e.
 $v_{\phi}\left(w,z\right)\,=v_{\phi}\left(w_{i},0\right)\,\left(w_{i}/w\right)$. 

The temperature of the wind 
($T_{\mathrm{dw}}$) is assumed to be isothermal. This is a reasonable
assumption because the studies of disc wind thermal structures by
e.g.~\citet{Safier:1993} and \citet{Garcia:2001} have shown that the
 wind is heated by ambipolar diffusion, and its temperature reaches $\sim
 10^{4}$\,K quickly after being launched. Further, their studies
 have shown that the wind temperature remains almost constant or only
 slowly changes as the wind propagates after the temperature reaches $\sim
 10^{4}$\,K.

\subsubsection{The K-band continuum-emitting ring}

\label{subsub:Ring-config}
Since the K-band continuum of the accretion discs of Herbig stars is
expected to originate mainly from the region near the dust sublimation
radius (e.g.~\citealt{Dullemond:2010}), we simulate this emission with
a geometrically thin ring on the equatorial plane, with its inner
radius $R_{\mathrm{ring}}$ and its width $\Delta R_{\mathrm{ring}}$
(Fig.~\ref{fig:model-config}). This is the same geometrical model used
to fit the continuum visibilities in
Section~\ref{sec:ring-model-fit}.  Again, we assume  
$\Delta R_{\mathrm{ring}}=0.2\, R_{\mathrm{ring}}$, as in MO05. The
ring is assumed to have a uniform temperature
$T_{\mathrm{ring}}=1500\,\Kelvin$, and to radiate as a blackbody.

\subsection{Modelling Br$\gamma$ line profiles}

\label{sub:model-line-prof}

Since computing the wavelength/velocity-dependent image maps is
 rather computationally expensive, we first concentrate on modelling
the Br$\gamma$ profile (top panel in Fig.~\ref{fig:amber-summary})
obtained with the AMBER instrument to find a reasonable set of model parameters 
which approximately matches the observed line profile. Then, we will
expand the model constraints to the rest of the observed
interferometric quantities (visibilities, differential and closure
phases) to further refine the model fit by computing the
wavelength/velocity-dependent image maps for each model
(Section~\ref{sub:model-vis}).

\subsubsection{Adopted stellar parameters}

\label{subsub:adopted-param}

The basic stellar parameters adopted for modelling the observed AMBER
data of our target HD~58647 are summarised in
Table~\ref{tab:stellar-param}.  We estimated the stellar luminosity
$L_*$ by fitting the SED of HD~58647 using the VOSA online SED
analyzing tool (\citealt{Bayo:2008}) using 
$T_{\mathrm{eff}}=10500$~K and $\log g_{*}=3.5$ (cgs)
(\citealt{Montesinos:2009}), the extinction $A_{V}=0.4$
(\citealt{Malfait:1998}) and the distance $d=318$\,pc
(VL07).  From this fit, we find
$L_{*}=412\,\Lsun$.  Using the evolutionary tracks of pre-main
sequence stars by \citet*{Siess:2000} with the stellar luminosity found
above and $\log g_{*}=3.5$, the stellar mass and radius of HD~58647
are estimated as $M_{*}=4.6\,\Msun$ and $R_{*}=6.2\,\Rsun$,
respectively.

The inclination angle $i$ of the K-band continuum-emitting ring was
estimated as $i\approx55^{\circ}$ in Section~\ref{sec:ring-model-fit}
(Table~\ref{tab:ring-fit-param})  
by fitting the continuum visibilities. Assuming the disc/ring normal vector coincides with the
rotation axis of HD~58647 and the speed at which the star rotates at
the equator ($v_{\mathrm{eq}}$) is same as $v$ in $v\sin i$, we obtain
$v_{\mathrm{eq}}=144\,\mathrm{km\, s^{-1}}$ (approximately $0.34$ of
the breakup velocity) using $v\sin i=118\,\kmps$
(\citealt{Mora:2001}). Consequently, the rotation period of HD~58647
is estimated as $P_{*}=2\pi R_{*}/v_{\mathrm{eq}}=2.2$~d.  We adopt
the mass-accretion rate $\dot{M}_{\mathrm{a}}=3.5\times
10^{-7}\,\MsunPerYear$ (\citealt{Brittain:2007}), which is estimated
from the CO line luminosity, in our magnetospheric accretion model.

%%%%%%%%%%%%%%%%%%%%%%%%%%%%%%%%%%%%%%%%%%%%%%%%%%%%%%%%%
\begin{table}

\caption{Adopted parameters of HD\,58647 models. \label{tab:stellar-param} }

\begin{center}
\begin{tabular}{ccccccc}
\hline 
$R_{*}$  & $M_{*}$  & $T_{\mathrm{eff}}$  & $P_{*}$ & $d$  & $\dot{M}_{\mathrm{a}}$ & $i$\tabularnewline
($\Rsun$)  & ($\Msun$)  & ($\mathrm{K}$)  & (d) & ($\mathrm{pc}$)  & ($\MsunPerYear$) & ($^{\circ}$)\tabularnewline
\hline 
$6.2$  & $4.6$  & $10500$  & $2.2$ & $318$  & $3.5\times10^{-7}$ & $55$\tabularnewline
\hline 
\end{tabular}

\end{center}

\end{table}
%%%%%%%%%%%%%%%%%%%%%%%%%%%%%%%%%%%%%%%%%%%%%%%%%%%%%%%%%

%%%%%%%%%%%%%%%%%%%%%%%%%%%%%%%%%%%%%%%%%%%%%%%%%%%%%%%%%
\begin{table*}

\caption{Main model parameters. \label{tab:model-param}}

\begin{tabular}{cccccccccccccc}
\hline 
 & \multicolumn{3}{c}{Magnetosphere} &  & \multicolumn{6}{c}{Disc
  wind} &  & \multicolumn{2}{c}{Continuum Ring}\tabularnewline
\cline{2-4} \cline{6-11} \cline{13-14} 
 & $R_{\mathrm{mi}}$ & $R_{\mathrm{mo}}$ & $T_{\mathrm{m}}$ &  & $\dot{M}_{\mathrm{dw}}$ & $R_{\mathrm{wi}}$ & $R_{\mathrm{wo}}$ & $D$ & $T_{\mathrm{dw}}$ & $\beta_{\mathrm{dw}}$ &  & $R_{\mathrm{ring}}$ & $\Delta R_{\mathrm{ring}}$\tabularnewline
Model  & ($R_{*}$) & ($R_{*}$) & ($\mathrm{K}$)  &  & ($\MsunPerYear$) & ($R_{*}$) & ($R_{*}$) & ($R_{*}$) & ($\mathrm{K}$) & $\cdots$ &  & ($R_{*}$) & ($R_{\mathrm{ring}}$)\tabularnewline
\hline 
A & $1.3$ & $1.7$ & $7500$ &  & $\cdots$ & $\cdots$ & $\cdots$ & $\cdots$ & $\cdots$ & $\cdots$ &  & $\cdots$ & $\cdots$\tabularnewline
B & $1.3$ & $1.7$ & $8000$ &  & $\cdots$ & $\cdots$ & $\cdots$ & $\cdots$ & $\cdots$ & $\cdots$ &  & $\cdots$ & $\cdots$\tabularnewline
C & $1.3$ & $1.7$ & $8500$ &  & $\cdots$ & $\cdots$ & $\cdots$ & $\cdots$ & $\cdots$ & $\cdots$ &  & $\cdots$ & $\cdots$\tabularnewline
D & $1.3$ & $1.7$ & $8000$ &  & $\cdots$ & $\cdots$ & $\cdots$ & $\cdots$ & $\cdots$ & $\cdots$ &  & $23.5$ & $0.2$\tabularnewline
E & $\cdots$ & $\cdots$ & $\cdots$ &  & $4.5\times10^{-8}$ & $1.7$ & $23.5$ & $100$ & $10^{4}$ & $2.0$ &  & $23.5$ & $0.2$\tabularnewline
F & $1.3$ & $1.7$ & $8000$ &  & $4.5\times10^{-8}$ & $1.7$ & $23.5$ & $100$ & $10^{4}$ & $2.0$ &  & $23.5$ & $0.2$\tabularnewline
\hline 
\end{tabular}

\end{table*}
%%%%%%%%%%%%%%%%%%%%%%%%%%%%%%%%%%%%%%%%%%%%%%%%%%%%%%%%%

\subsubsection{Br$\gamma$ profiles from magnetospheric accretion models}

\label{subsub:line-MA-model}

Here, we briefly examine the effect of magnetospheric accretion
(Section~\ref{subsub:MA-config}) on the formation of the hydrogen
recombination line Br$\gamma$.  Using the stellar
parameters above, the corotation radius ($R_{\mathrm{cr}}$) of
HD~58647 is estimated as 
$R_{\mathrm{cr}}=\left(GM_{*}\right)^{1/3}\left(P_{*}/2\pi\right)^{2/3}=1.9\,R_{*}$.
The extent of the magnetosphere is assumed to be slightly smaller than
$R_{\mathrm{cr}}$; hence, we set the inner and outer radii of the
magnetospheric accretion funnel as $R_{\mathrm{mi}}=1.3\, R_{*}$ and
$R_{\mathrm{mo}}=1.7\, R_{*}$, respectively. 
These radii (in units of the
stellar radius) are considerably smaller than those of classical
T~Tauri stars (e.g.~\citealt{Koenigl:1991};
\citealt{Muzerolle:2004}) which typically has larger rotation periods,
i.e. 7--10\,d (e.g.~\citealt{Herbst:2002}).  The smaller magnetosphere
size in HD~58647 leads to smaller volumes for the accretion
funnels. Hence, the hydrogen line emission is also expected to be 
much weaker. \cite{Muzerolle:2004} also found that the emission lines
from a small and rotating magnetosphere were very weak in their
Balmer line models for the Herbig Ae star UX~Ori (see their Fig.~3). 

\begin{figure}

\begin{center}

\includegraphics[clip,width=0.4\textwidth]{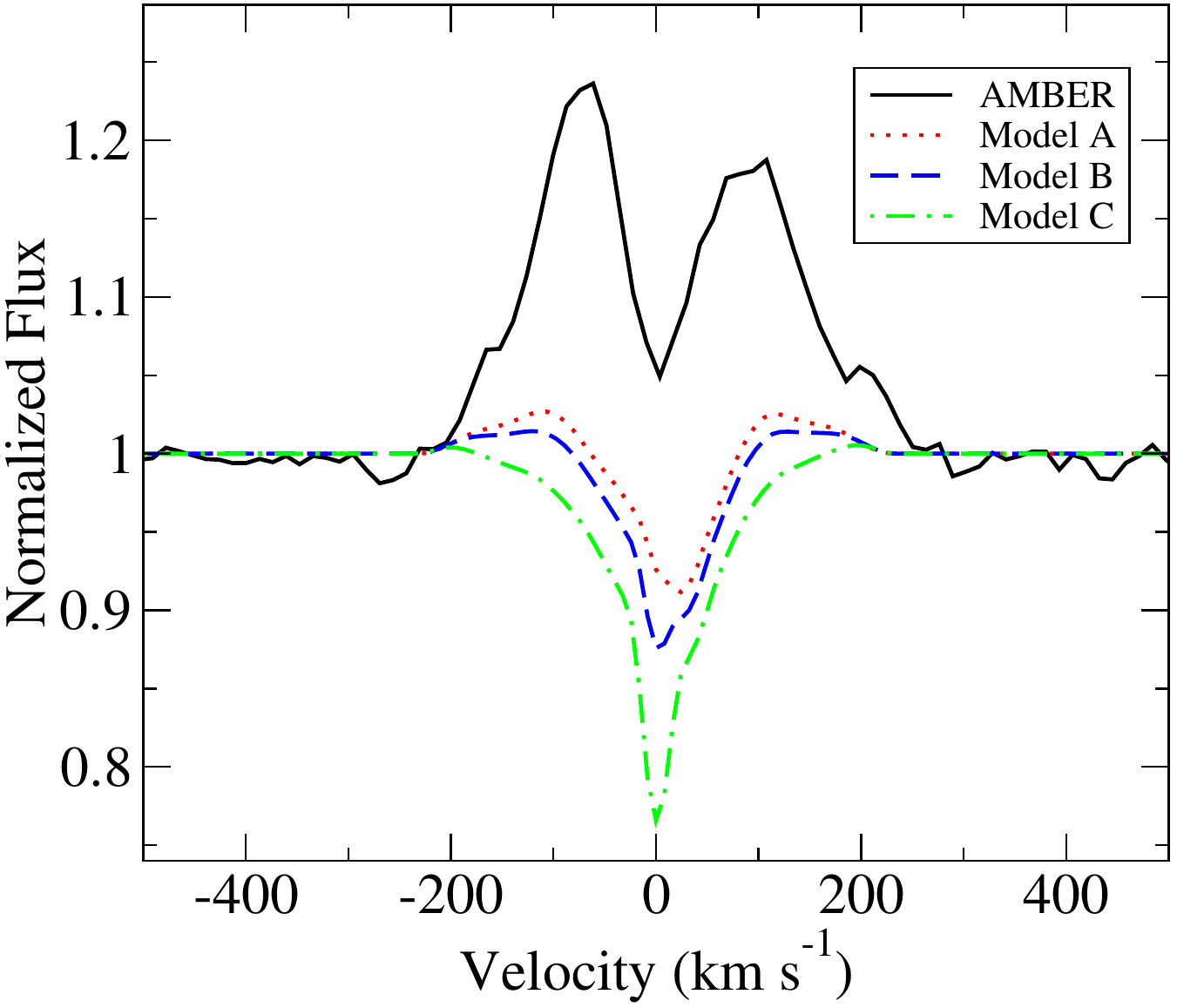}
\caption{Comparison of the observed Br$\gamma$ line profile (solid)
  with the line profiles computed at the inclination angle
  $i=55^{\circ}$ with a rotating magnetosphere with different
  temperatures: $T_{\mathrm{m}} =$ 7500 (Model~A, dot), 8000 (Model~B,
  dash) and 8500\,K (Model~C, dash-dot). The models use the
  magnetosphere with $R_{\mathrm{mi}}=1.3\,\Rsun$,
  $R_{\mathrm{mo}}=1.7\,\Rsun$ and
  $\dot{M}_{\mathrm{a}}=3.5\times10^{-7}\,\MsunPerYear$
  (\citealt{Brittain:2007}).  All the models show a rather strong
  central absorption component. Only the models with $T_{\mathrm{m}} <
  8000$\,K show very weak emission components.  For clarity, the
  continuum emission from the ring in the accretion disc and
  the photospheric absorption component are not
  included here. \label{fig:ma-tm-effect} }

\end{center}

\end{figure}

Fig.~\ref{fig:ma-tm-effect} shows a comparison of the observed
Br$\gamma$ profile from the AMBER observation
(Section~\ref{sec:observations}) with the model profiles computed using the
parameters for HD~58647 (Table~\ref{tab:stellar-param}).  The models
are computed for 3 different magnetospheric temperatures:
$T_{\mathrm{m}}= 7500$, 8000 and 8500\,K  (Models~A, B and C in
Table~\ref{tab:model-param}, respectively). The continuum emission
from the ring are omitted here so that the emission component from
the magnetosphere can be seen more easily.  As expected, the lines are
rather weak, and are mainly in absorption 
except for very small emission components around the velocities $v=\pm
100\,\kmps$ in the lower temperature models (Models~A and B). No
significant emission above the continuum level is seen 
for models with $T_{\mathrm{m}}>8500\,\Kelvin$. Similar line
shapes and strength are also found in the small magnetosphere model of
\cite{Muzerolle:2004} (their Fig.~3). 

The models clearly disagree with the observed line profiles in their
emission strengths, i.e.\,the emission from the magnetosphere is much
weaker than the observed one. If the ring continuum emission is
included in the models, the emission components in the models will
appear even weaker. As suggested by \cite{Muzerolle:2004}, one way to
increase the emission strength is to increase the size of
magnetosphere; however, this will be in contradiction with the stellar
parameters estimated earlier. It is most likely that an
additional gas flow component is involved in the formation of the
Br$\gamma$ emission line. This possibility will be explored next.

%%%%%%%%%%%%%%%%%%%%%%%%%%%%%%%%%%%%%%%%%%%%%%%%%%%%%%%%%
%%%%%%%%%%%%%%%%%%%%%%%%%%%%%%%%%%%%%%%%%%%%%%%%%%%%%%%%%

%%%%%%%%%%%%%%%%%%%%%%%%%%%%%%%%%%%%%%%%%%%%%%%%%%%%%%%%%
\begin{table}

\caption{Ranges of the disc wind and continuum-emitting ring model
  parameters explored. \label{tab:dw-param-range} }

\begin{center}
\begin{tabular}{ccccc}
\hline 
$\dot{M}_{\mathrm{dw}}$ & $D$ & $T_{\mathrm{dw}}$ & $\beta_{\mathrm{dw}}$ & $R_{\mathrm{ring}}$\tabularnewline
($10^{-8}\,\MsunPerYear$) & ($R_{*}$) & ($\mathrm{K}$) & $\cdots$ & ($R_{*}$)\tabularnewline
\hline 
3.0--6.0 & 50--200 & 9000--11000 & 1.5--2.5 & 21--27\tabularnewline
\hline 
\end{tabular}
\par\end{center}

\end{table}
%%%%%%%%%%%%%%%%%%%%%%%%%%%%%%%%%%%%%%%%%%%%%%%%%%%%%%%%%

%%%%%%%%%%%%%%%%%%%%%%%%%%%%%%%%%%%%%%%%%%%%%%%%%%%%%%%%%
\begin{figure}

\begin{center}

\includegraphics[clip,width=0.4\textwidth]{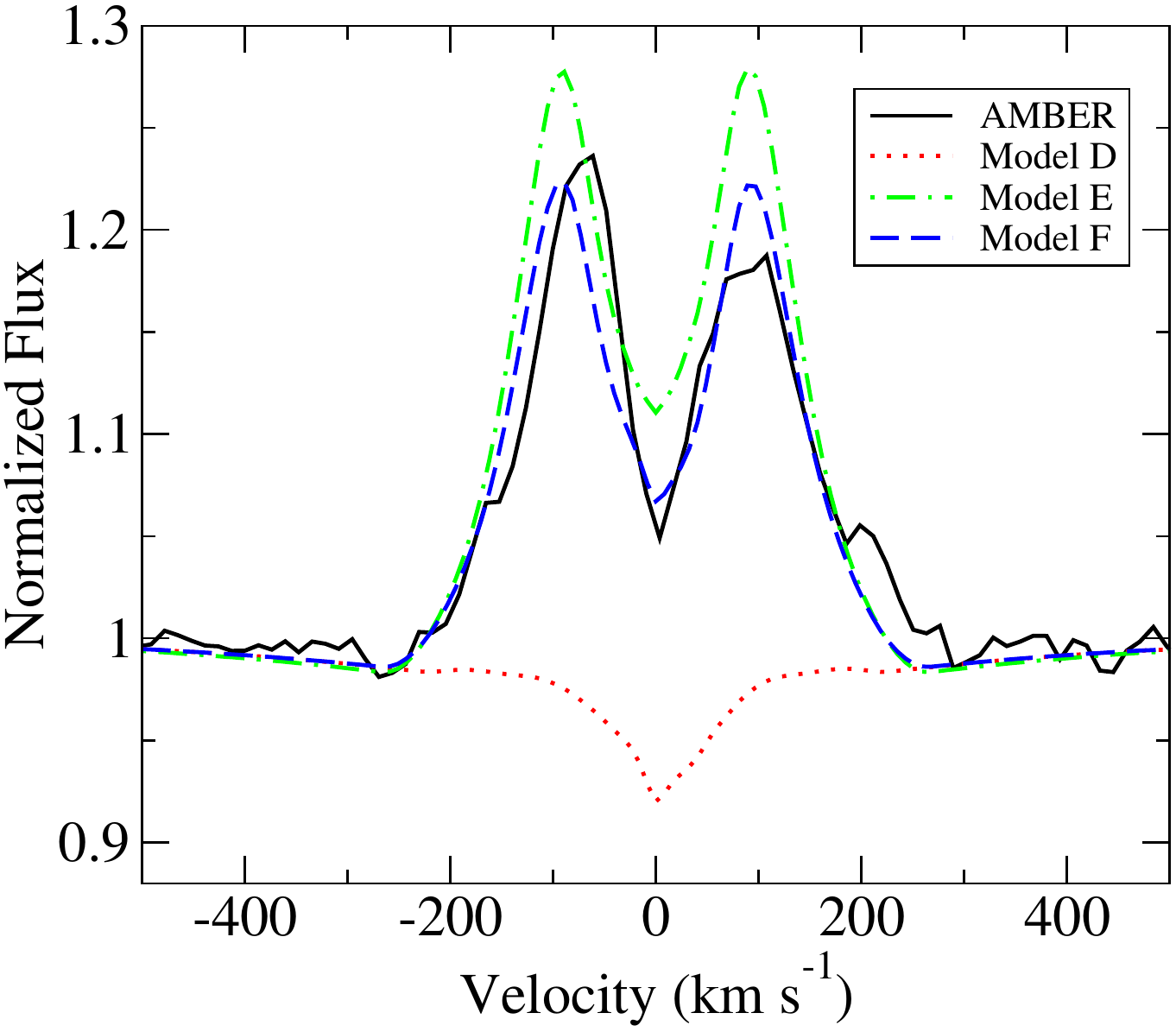}

\caption{Comparison of the Br$\gamma$ line profiles computed with
  (1)~a magnetosphere only (Model~D, dotted), (2)~a disc wind only
  (Model~E, dash-dot) and (3)~a magnetosphere and a disc wind
  (Model~F, dash). The models are computed at $i=55^{\circ}$ and with
  the continuum-emitting ring and photospheric absorption
  component.  The main model parameters used are  
  summarised in Table~\ref{tab:model-param}. The Br$\gamma$ profile
  from the AMBER observation (solid) is also shown for a comparison.
  The line profile from the magnetosphere only model (Model~D) is
  mainly in absorption, with relatively strong absorption near the
  line centre.  The model with the magnetosphere and disc wind
  (Model~F) fit the observation reasonably well. The central
  absorption dip is enhanced by the absorption in the magnetosphere,
  and the corresponding match with the observation is slightly better
  than the model with the disc wind alone (Model~E).  \label{fig:compare-ma-dw}}

\end{center}

\end{figure}
%%%%%%%%%%%%%%%%%%%%%%%%%%%%%%%%%%%%%%%%%%%%%%%%%%%%%%%%%

\subsubsection{Br$\gamma$  profiles from disc wind models }

\label{subsub:line-DW-model} 

To bring the line strength of the model Br$\gamma$ to a level
consistent with the one detected in the AMBER spectra, we now
consider an additional flow component,  
namely the disc wind, as briefly described in
Section~\ref{subsub:DW-config}.  We have explored various combinations
of the disc wind model parameters (i.e.~$\dot{M}_{\mathrm{dw}}$,
$D$, $T_{\mathrm{dw}}$ and $\beta_{\mathrm{dw}}$ as
in Section~\ref{subsub:DW-config}; see also
Fig.~\ref{fig:model-config}) and the continuum-emitting ring size
($R_{\mathrm{ring}}$ as in Section~\ref{subsub:Ring-config}) to fit
the observed Br$\gamma$ from  the AMBER observation. The range of the
model parameters explored are summarised in Table~\ref{tab:dw-param-range}. 
In this analysis, we set the inner and outer wind radii
($R_{\mathrm{wi}}$ and $R_{\mathrm{wo}}$) to coincide with the outer
radius of the magnetosphere $R_{\mathrm{mo}}$ and the ring radius
($R_{\mathrm{ring}}$), respectively. $R_{\mathrm{wi}}$ is set to the
magnetospheric radius because a magnetosphere would set the inner
radius of the accretion disc where the disc wind mass-loss flux would
be highest (e.g.~\citealt{krasnopolsky:2003}), and because a wind
could arise in the disc-magnetosphere interaction region
(e.g.~\citealt{shu:1994,Shu:1995}; \citealt{Romanova:2009}). The outer
wind launching radius $R_{\mathrm{wo}}$ is set to
$R_{\mathrm{ring}}$ because the size of Br$\gamma$ emission was found 
to be smaller than the size of the K-band continuum-emitting region in
Section~\ref{sec:observations}, i.e. the visibility level increases in the line (by
$\sim 0.1$) at the longest projected baseline of the VLTI-AMBER
observation (Fig.~\ref{fig:amber-summary}). Hence, no significant wind
emission is expected to be seen at radial 
distances beyond $R_{\mathrm{ring}}$.

Fig.~\ref{fig:compare-ma-dw} shows that our model with a magnetosphere
plus disc wind (Model~F) can reasonably
reproduce the observed Br$\gamma$ profile. The figure also shows the
models computed with a magnetosphere only (Model~D) and a disc wind
only (Model~E) to demonstrate the contributions from each
component. The corresponding model parameters are shown in
Table~\ref{tab:model-param}.  As one can see from the figure, the
emission line is much stronger when the 
emission from the disc wind is included in the model. The match
between the observation and Model~F is quite good. As seen in the
previous section, the line profile from the magnetosphere only model
(Model~D) is mainly in absorption, and the emission from the disc
wind is dominating the line.  The central absorption dip is
enhanced by the absorption in the magnetosphere, and the corresponding
match with the observation is slightly better for Model~F, compared to the
model with the disc wind alone (Model~E).
In Model~F, the ratio of the mass-loss rate in the wind to the
mass-accretion rate ($\mu =
\dot{M}_{\mathrm{dw}}/\dot{M}_{\mathrm{a}}$) is about 0.13.

A prominent characteristics in the model profiles is their
double-peaked appearance. Since these lines are mainly formed near the
base of the disc wind where the Keplerian rotation of the wind is dominating
over the poloidal motion, the double-peak profiles naturally occurs
when the disc wind is viewed edge-on, i.e.,
when a system has a mid to high inclination angle ($i$). The extent of 
the line profile can be also explained by the Keplerian velocity
($V_{\mathrm{k}}$) of the disc. At the inner radius of the disc wind
($R_{\mathrm{wi}}=1.7\,R_{*}$), $V_{\mathrm{k}}=289\,\kmps$. The
corresponding projected velocity is
$V_{\mathrm{k}}\sin55^{\circ}=237\,\kmps$, which is very similar to
the extent of the observed Br$\gamma$ profile. This also indicates
that our choice of the inner radius of the disc wind, which coincides
with the outer radius of the magnetosphere, is reasonable.

%%%%%%%%%%%%%%%%%%%%%%%%%%%%%%%%%%%%%%%%%%%%%%%%%%%%%%%%%
\begin{figure}

\begin{center}

\begin{tabular}{c}
   \hspace{-1.0cm}
   \vspace{+0.3cm}
   \includegraphics[clip,width=0.35\textwidth]{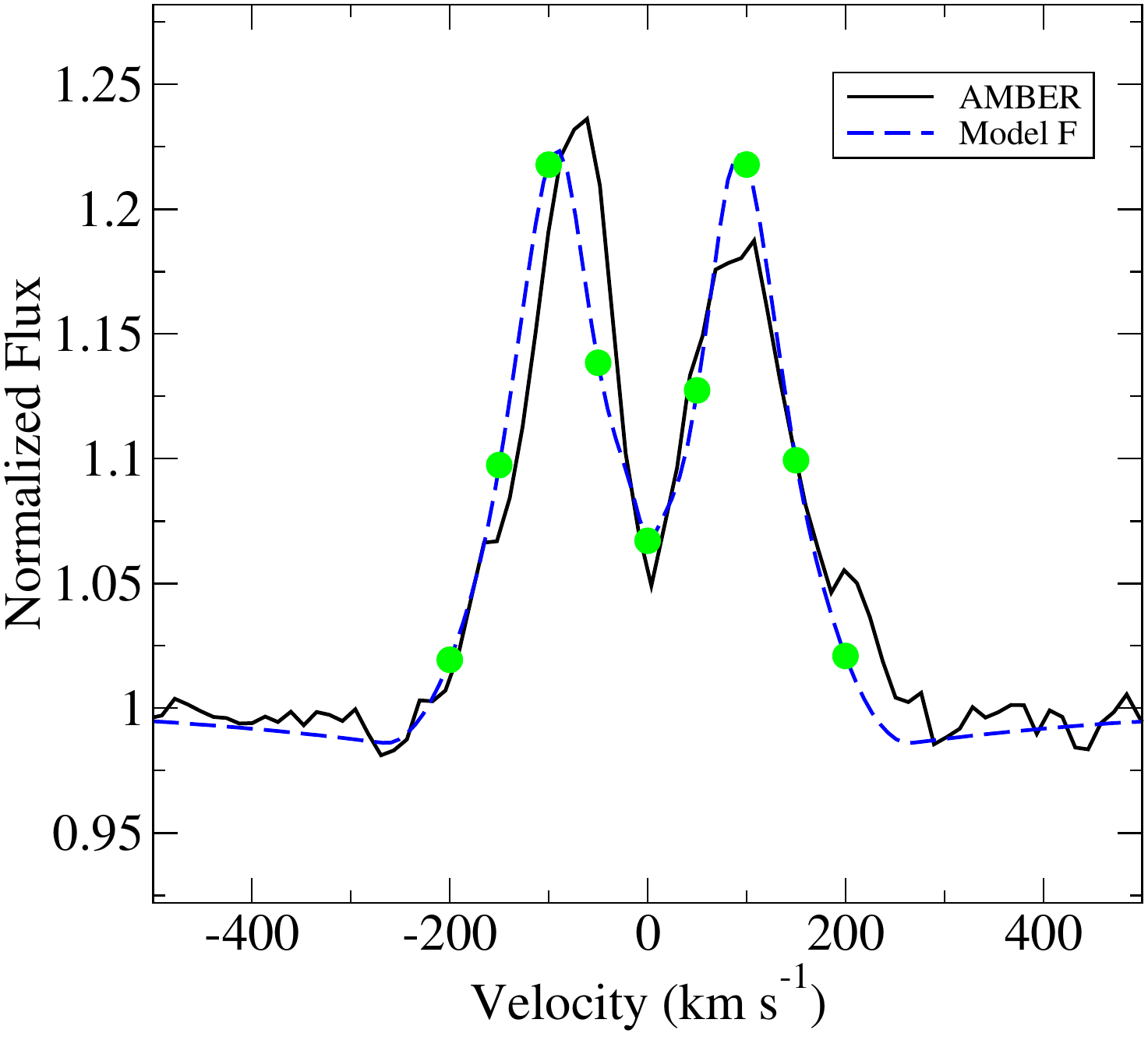}\tabularnewline
   \includegraphics[clip,width=0.48\textwidth]{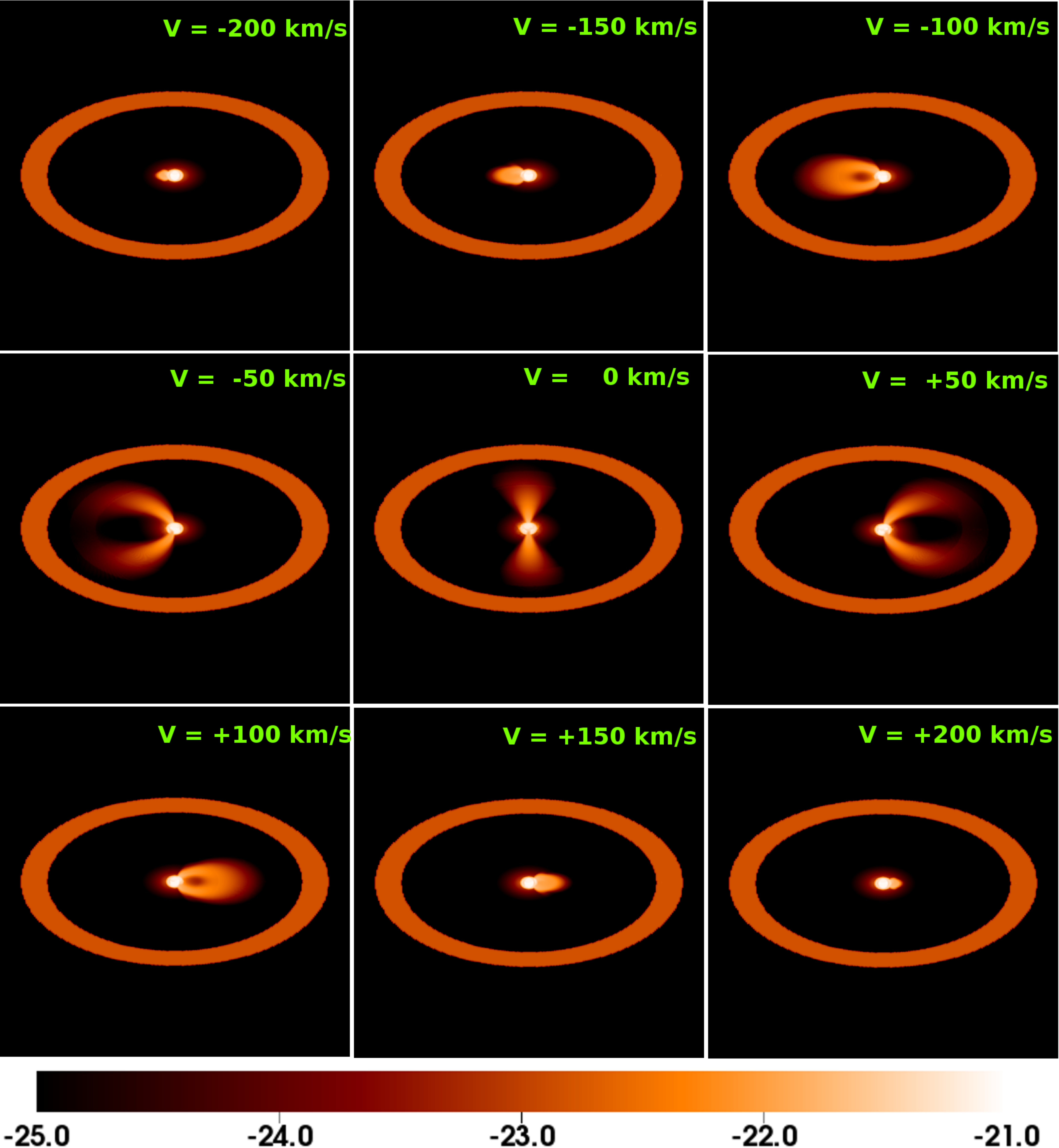}\tabularnewline
\end{tabular}

\caption{Upper panel: the model Br$\gamma$ profile (dash) computed at
  $i=55^{\circ}$ with the magnetosphere and disc wind (Model~F in
  Table~\ref{tab:model-param}) is compared with the AMBER observation
  (solid). The velocity channels at which the emission maps (lower
  panel) are computed are indicated by filled circles. The
  corresponding emission maps (in logarithmic scale and in arbitrary
  units) at 9 different velocity channels are shown in the lower
  panel. The image sizes are 1.89 $\times$ 1.89 $\mathrm{au}^{2}$ or
  5.94 $\times$ 5.94~$\mathrm{mas}^2$, assuming the object is at 318~pc
  (VL07). The wavelength/velocity dependent
  images shown here are used to compute the interferometric quantities
  such as visibilities, differential phases and closure phases as a
  function of wavelength/velocity. \label{fig:vel-image}}

\end{center}

\end{figure}
%%%%%%%%%%%%%%%%%%%%%%%%%%%%%%%%%%%%%%%%%%%%%%%%%%%%%%%%%

%%%%%%%%%%%%%%%%%%%%%%%%%%%%%%%%%%%%%%%%%%%%%%%%%%%%%%%%%
\begin{figure*}

\begin{center}

\begin{tabular}{ccc}

\includegraphics[clip,width=0.31\textwidth]{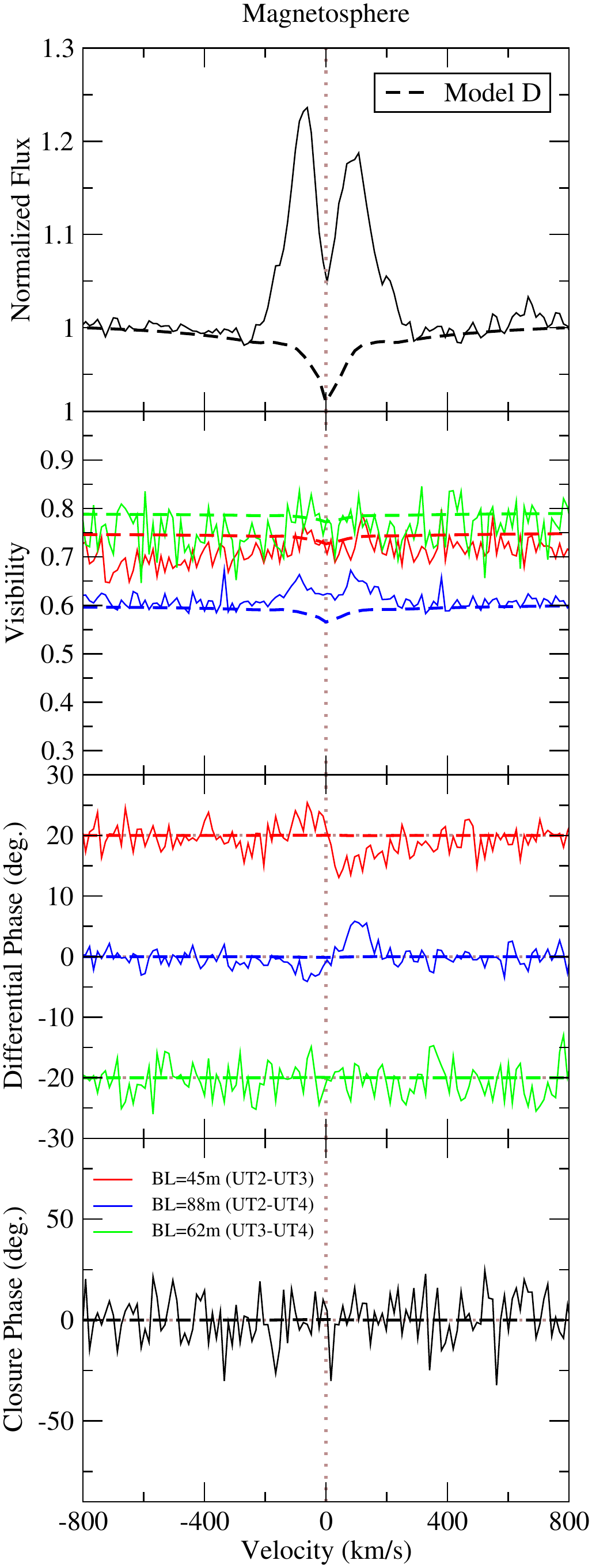} & 
\includegraphics[clip,width=0.31\textwidth]{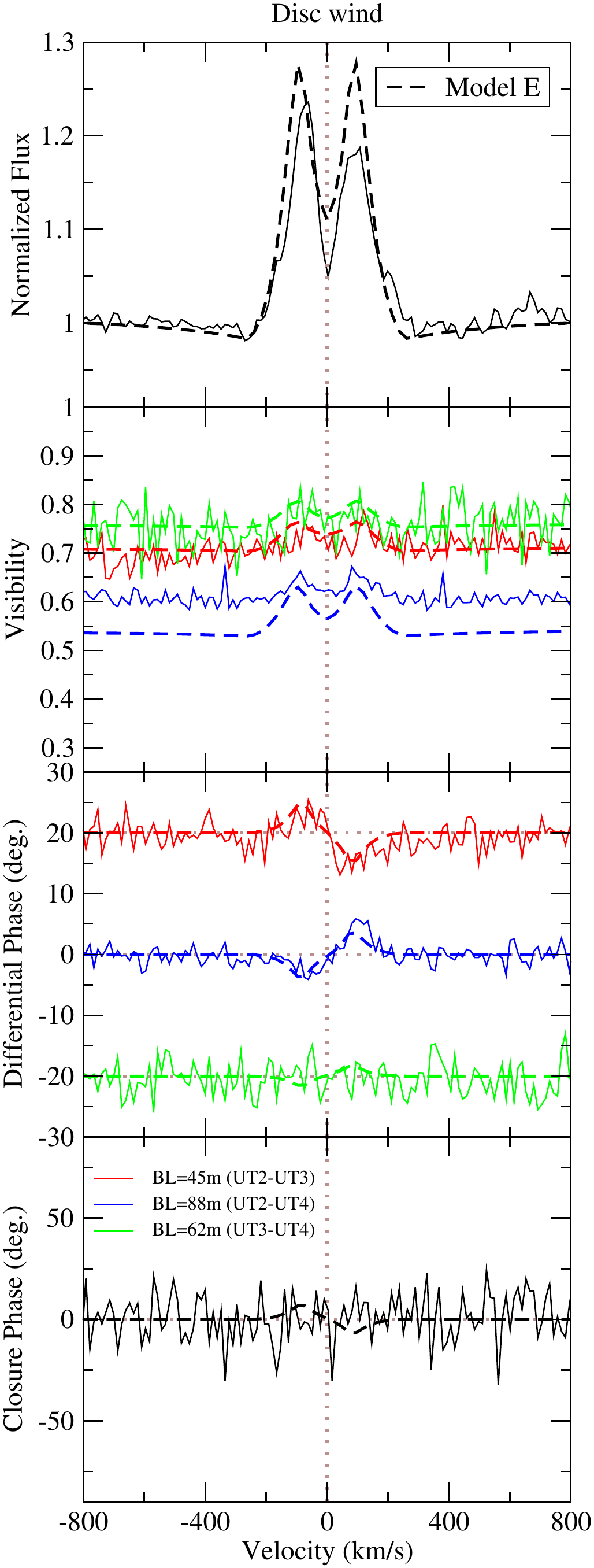} & 
\includegraphics[clip,width=0.31\textwidth]{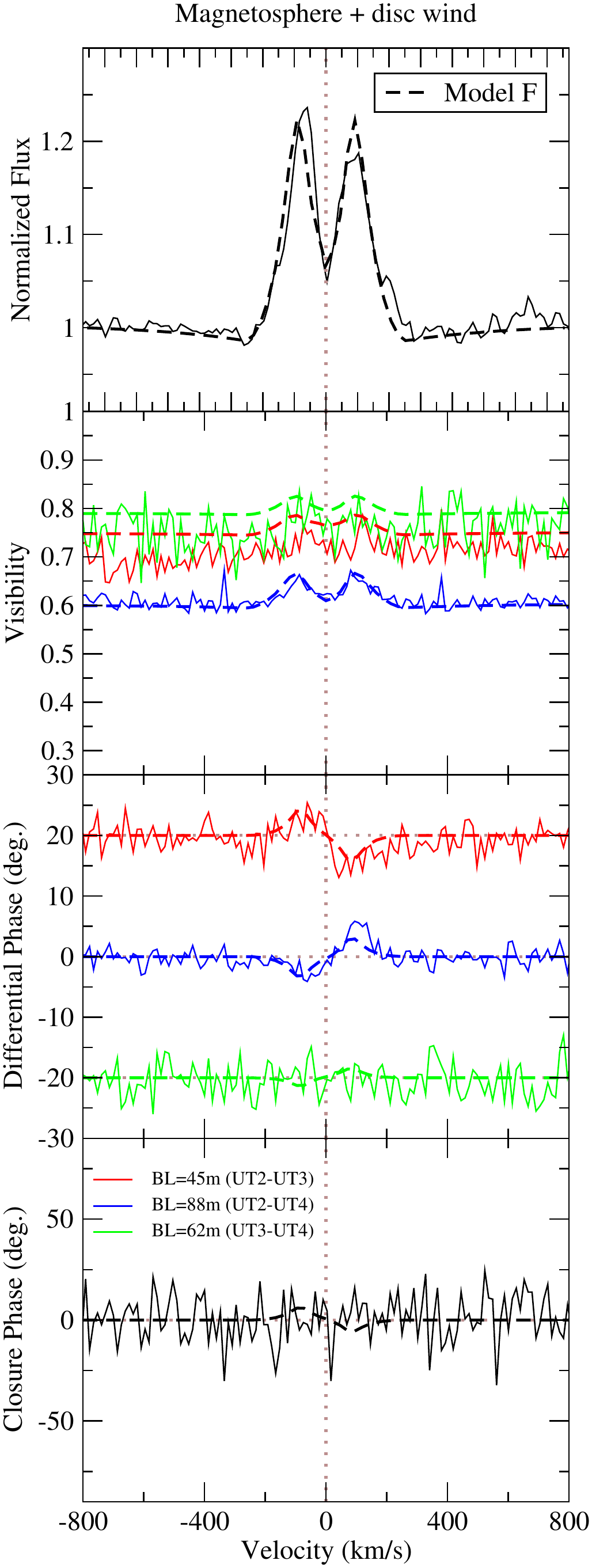}\tabularnewline 
\end{tabular}

\caption{Summary of the model fits to the high-spectral-resolution
  (R=12000) AMBER observation shown in
  Fig.~\ref{fig:amber-summary}. The wavelength/velocity-dependent
   normalised flux, visibilities,
  differential phases and closure phases are shown from top to
  bottom. In all the panels, the observed values are
  shown in solid and the models are in dashed lines. The projected
  VLTI baselines are 45.4\,m (red), 87.9\,m (blue) and
  62.2\,m (green), and their corresponding position angles are
  $37^{\circ}$, $259^{\circ}$ and $288^{\circ}$, respectively. The
  differential phases for the first (red) and third (green) baselines
  are vertically shifted by $+20^{\circ}$ and $-20^{\circ}$,
  respectively, for clarity. See Fig.~\ref{fig:amber-summary} for the
  uncertainties in the absolute calibration of the visibilities. The figure
  compares three cases: (1)~the model with magnetospheric accretion alone
  (Model~D, left panels), (2)~the model with the disc wind alone
  (Model~E, middle panels), and (3)~the model with a magnetosphere and
  a disc wind (Model~F, right panels).  All the models shown here use
  the inclination angle $i=55^{\circ}$ and the system axis position angle
  $\mathrm{PA} = 85^{\circ}$. Model~F (right panels) fits all the
  observed interferometric quantities well. While Model~F shows the
  best fit to the observations here, the disc wind only
  model (Model E) could be improved to fit the observations
  (in particular the visibility levels for the longest baseline) by
  using a smaller continuum ring radius and by adjusting other model
  parameters accordingly. \label{fig:amber-model-best} }

\end{center}

\end{figure*}
%%%%%%%%%%%%%%%%%%%%%%%%%%%%%%%%%%%%%%%%%%%%%%%%%%%%%%%%%

\subsection{Modelling spectro-interferometric data}
\label{sub:model-vis}

To examine whether the magnetosphere plus disc wind model used to fit
the observed Br$\gamma$ line profiles can also fit the other
interferometric data obtained by VLTI-AMBER
(Fig.~\ref{fig:amber-summary}), we compute the intensity maps of the
line-emitting regions plus the ring continuum-emitting region using
the same radiative transfer model as in
Section~\ref{subsub:line-DW-model}.  The model intensity maps allow us
to compute the wavelength-dependent visibilities, differential and
closure phases, and to compare them directly with the observation.
As in the geometrical ring fit for the LR VLTI-AMBER continuum
visibilities in Section~\ref{sec:ring-model-fit}, the position angle (PA)
of the system is measure from north to east, i.e. the disc/ring normal axis
coincides with the north direction when PA is zero, and PA 
increase as the disc/ring normal axis rotates towards east
(cf. Fig.~\ref{fig:uv-coverage}). 

Examples of the intensity maps computed for Model~F
(Table~\ref{tab:model-param}), with the velocity interval $\Delta v =
\,50 \kmps$ between -200 and 200\,$\kmps$, are shown in
Fig.~\ref{fig:vel-image}. Here, the images are computed with
$\mathrm{PA}=0\degr$ for a demonstration. The figure also shows the corresponding
model Br$\gamma$ profile marked with the same velocity channels used
for the images. As expected, the shape of the ring continuum
emission does not change across the line, and the emission is
uniform. On the other hand, the wind emission size and shape greatly
depend on the wavelength/radial velocity. The dependency is consistent with
the motion of the gas in the disc wind. The wind base is
rotating at the local Keplerian velocities of the accretion disc
(between 289 and 78\,$\kmps$ from $R_{\mathrm{wi}}=1.7\,R_{*}$ to
$R_{\mathrm{wo}}=23.5\,R_{*}$).   As one can see from the figure, the
wind on the left-hand side is approaching us, and that on the
right-hand side is moving away from us due to the rotational motion of
the wind.  If the motion of the Br$\gamma$-emitting gas is dominated by the poloidal
component, the images would rather have a left-right symmetry. However,
this is not the case in our intensity maps. This also indicates that
the Br$\gamma$ emission mainly occurs near the base of the wind where
the rotational motion dominates.  The radial extent of the Br$\gamma$
emission in the disc wind is estimated as $\sim$0.5~au from the
intensity maps (with the velocity channels $\pm 50\,\kmps$ which show
the maximum radial extent of the line emission).

By using the emission maps and scaling their sizes according to the
distance to HD~58647 (318~pc, VL07), we have
computed the wavelength-dependent visibilities, differential and
closure phases for Models~D, E and F. Here, the maps are computed
using the same velocity interval as in the AMBER observations, i.e. $\Delta v \approx
13\, \kmps$. 
The computed images are rotated
by the PA of the system, which is assumed to be a
free parameter. Since the continuum-emitting ring and the disc wind are not spherical
symmetric, the visibilities levels, differential and closure phases
are sensitive to the PA of the system. Hence, these interferometric data
are used as additional constraints for the model fitting procedure
with the PA as an additional parameter.  
The results are summarised in Fig.~\ref{fig:amber-model-best}. The figure
shows the models computed with the inclination angle $i=55^{\circ}$, 
the system position angle $\mathrm{PA} = 85^{\circ}$ and the
continuum-emitting ring radius $R_{\mathrm{Ring}}=23.5\,R_{*}$ for the
3 different cases: (1)~the model with magnetospheric accretion only 
(Model~D), (2)~the model with the disc wind only (Model~E), and
(3)~the model with a magnetosphere and disc wind (Model~F). 

In the first case (Model~D: with magnetosphere; see
Table~\ref{tab:model-param}), the agreement of the continuum 
visibilities with the AMBER observation is excellent, but large
disagreements in the line profile shape (as in
Sections~\ref{subsub:line-MA-model} and \ref{subsub:line-DW-model}) and the
differential phases in the line can be seen. 
The angular size of the magnetosphere is about 0.15\,mas; hence, it is
unresolved by the interferometer. There is a slight indication
for a depression in the visibility level at the line centre due to the
magnetospheric absorption. The photo-centre shifts due to the
presence of the magnetosphere are very small on all baselines; hence,
there is no change in the differential phases within the line.  This
 clearly disagrees with the observed differential phases which show
 S-shape signals across the line for two baselines. Similarly, the
 model closure phases are essentially zero across the line due to the
small angular size of the magnetosphere, which is consistent with
the observations within in their large error bars ($\sim \pm 20^{\circ}$ as
in Fig.~\ref{fig:amber-summary}). 

In the second case (Model~E: with a disc wind; see
Table~\ref{tab:model-param}), the agreement of the differential phases
with the observation is excellent. This indicates that the disc wind
model used here is reasonable. Main disagreements are seen in the line
strength and the visibility levels for the longest project
baseline (88\,m: UT2-UT4). The model Br$\gamma$ line profile is slightly too strong
compared to the observed one, and the continuum visibility level for
the longest baseline in the model is too low compared to the
observation.  If we adopt a smaller ring size, the visibility levels
for the longest baseline would increase, but the line strength (with
respect to the continuum) would also increase.  Hence, the disagreement
in the line profiles would also increase unless we decrease the line
strength by decreasing, for example, the disc wind mass-loss rate
($\dot{M_{\mathrm{dw}}}$). 
The situation can be improved if there is an additional 
continuum emission component which is more compact than the ring
continuum emission. For example, the continuum emission from the
magnetosphere as in the first case (Model D) will make the line weaker
and make the visibility slightly larger.  A small S-shape variation in
closure phases across the line is seen in the model due to the
non-spherical symmetric nature of the disc wind emission. The
amplitude of the variation is relatively small ($\sim 7^{\circ}$), and
it does not disagree with the data which have rather large error bars
($\sim \pm 20^{\circ}$ as in Fig.~\ref{fig:amber-summary}).

In the third case, the model (Model~F: with a magnetosphere and disc
wind; see Table~\ref{tab:model-param}) fits all the observed
interferometric quantities, including the line profile, very well.
This is our best case of the three models. In particular, the model
simultaneously reproduces the following characteristics in the
observed interferometric data: (1)~the double-peak Br$\gamma$ profile,
(1)~the double-peak variation of the visibilities across the line
(especially for the longest project baseline), (3)~the S-shape
variations of the differential phases across the line, and (4)~the
small amplitude in the variation of the closure phases across the
line.  The locations of the peaks in the visibilities curves are
approximately $\pm 50\,\kmps$, which corresponds to the two peaks in
the line profile. The line emission maps shown earlier
(Fig.~\ref{fig:vel-image}) also suggests that the extent of the line
emission from the disc wind is largest at these velocities.

In summary, our radiative transfer model (Model~F in
Table~\ref{tab:model-param}), which uses a combination of a 
small magnetosphere and a disc wind with its inner radius
at the outer radius of the magnetosphere, is in good agreement with
the observed interferometric data for HD~58647. The extent of the
line-emitting disc wind region (the outer radius 
of the intensity distribution $\sim0.5$~au as in
Fig.~\ref{fig:vel-image}) is slightly smaller 
than the K-band continuum-emitting ring size
($R_{\mathrm{ring}}=23.5\,R_{*}$ or 0.68~au) which approximately 
corresponds to the dust sublimation radius of the system.

\section{Discussion }

\label{sec:discussion}

\subsection{Continuum-emitting regions}
\label{sub:discuss-cont-region}

In Section~\ref{sec:ring-model-fit}, we used a simple geometrical model
(an elongated ring) and the observed continuum
visibilities to estimate the size of the K-band continuum-emitting
region. The angular radius of the ring emission was estimated as
$R_{\mathrm{ring}}=2.0$~mas  or equivalently
0.64~au at the distance of 318~pc (VL07)
(Table~\ref{tab:ring-fit-param}). This radius is 
slightly larger than the value obtained by MO05, i.e.\,$1.47$~mas.
However, their radius determination does not include the effect of ring
elongation/inclination because the $uv$-coverage of the Keck
Interferometer in MO05 provides almost no two-dimensional
information.  Further, the ring radius estimated in
Section~\ref{sec:ring-model-fit} 
(0.64~au) is slightly smaller than the value used in the radiative
transfer models (Table~\ref{tab:model-param}),
i.e.\,$R_{\mathrm{ring}}=0.68$~au. 
The larger continuum radius is used in the radiative transfer model to
reproduced the observed visibilities because of the additional compact
continuum emission from the magnetosphere, which is not 
included in the geometric model in Section~\ref{sec:ring-model-fit}. 

Assuming a dust sublimation temperature of 1500~K and the stellar luminosity
of  $L_{*}=412\,\Lsun$ 
(Section~\ref{subsub:adopted-param}), the size-luminosity
relationship in MO05 provides a dust sublimation radius
$R_{\mathrm{s}}$ of $0.64$~au. 
This is very similar to the ring radius
$R_{\mathrm{ring}}=0.68$~au used in the radiative transfer model
(Model~F), which fits our AMBER continuum visibility observation
(Fig.~\ref{fig:amber-model-best}).

\subsection{Disc wind launching regions}
\label{sub:discuss-line-region}
\begin{figure}

\begin{center}

\includegraphics[clip,width=0.48\textwidth]{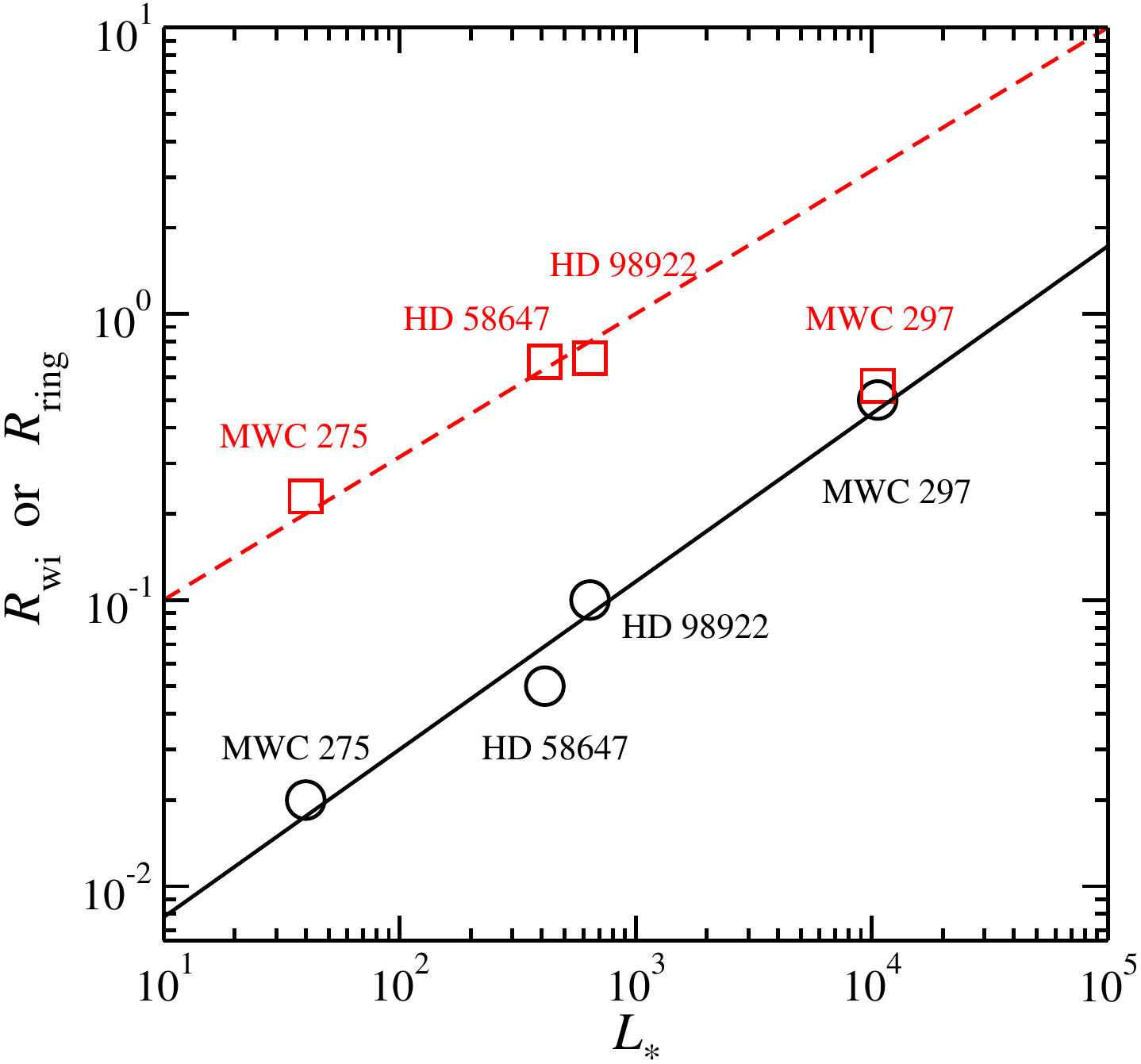}

\caption{Comparison of the disc wind inner radii
  ($R_{\mathrm{wi}}$) for the Herbig Ae/Be stars observed with the
  VLTI-AMBER in the high spectral resolution mode: MWC~275
  (\citealt{GarciaLopez:2015}), HD~58647 (this work),  HD~98922
  (\citealt{Caratti:2015}) and MWC~297
  (\citealt{Weigelt:2011}).  The inner radii of the disc wind used in 
  radiative transfer models (circle) are shown as a function of
  stellar luminosity ($L_{*}$).  
  The slope of the power-law fit (solid) for the
  disc wind inner radii is $0.59(\pm0.07)$. The radius of the K-band
  continuum-emitting ring ($R_{\mathrm{ring}}$, square) and the expected dust
  sublimation radii from MO05 (dash, assuming the dust sublimation temperature
  of 1500\,K) are also shown for a comparison. The units for
  the radii ($R_{\mathrm{wi}}$ and $R_{\mathrm{ring}}$) and the
  luminosity ($L_{*}$) are in au and $\Lsun$, respectively.  See
  Table~\ref{tab:amber-hr-sizes} for the numerical values used for
  this plot. \label{fig:compare-sizes} }

\end{center}

\end{figure}

According to the disc wind model (Model~F), which fits the
spectro-interferometric observation in Section~\ref{sub:model-vis}
well, the inner radius of the disc-wind launching region ($R_\mathrm{wi}$)
is only 0.05~au, which is located just outside of the 
magnetosphere.  This radius is much smaller than the dust sublimation radius
($R_\mathrm{s}$), which is approximately equal to the radius of the
continuum-emitting geometrical ring $R_{\mathrm{ring}}=0.68$~au, as
shown in the previous section. Next, we compare the size of the inner
radius of the disc-wind launching region of our HD~58647 model with
those found in earlier studies of other Herbig Ae/Be stars, in which
the VLTI-AMBER in the high spectral 
resolution mode and radiative transfer models for
Br$\gamma$ were used. Only three such studies are available in the
literature: the Herbig Be star MWC~297 (\citealt{Weigelt:2011}), the
Herbig Ae star MWC~275 (\citealt{GarciaLopez:2015}) and the Herbig Be
star HD~98922 (\citealt{Caratti:2015}). These studies used similar
kinematic disc wind models to probe the wind-launching region of these
stars. Fig.~\ref{fig:compare-sizes}
shows the inner radii of the disc wind launching regions plotted as a
function of stellar luminosity (see also
Table~\ref{tab:amber-hr-sizes}).\footnote{\citet{Benisty:2010},
  \citet{Kraus:2012b}, 
  \citet{Ellerbroek:2015} and \citet{Mendigutia:2015} presented the high spectral resolution 
VLTI-AMBER observations of Z~CMa, V921 Sco, HD~50138 and HD~100564, respectively;
however, their observations were not analysed with a radiative
transfer model with a disc wind. Hence, we excluded their
results from this analysis.} The figure shows that the inner radius of
the wind launching region 
increases as the luminosity increases.  By fitting the data with a
power-law, we find the following relation: 
\begin{equation}
   \left(\frac{R_{\mathrm{wi}}}{1\,\mathrm{au}}\right)
    =a\left(\frac{L_{*}}{1\,\Lsun}\right)^{b}
\label{eq:dw-inner-radius}
\end{equation}
where $a=2.0(^{+1.1}_{-0.7}) \times 10^{-3}$ and $b=0.59(\pm0.07)$.
As in Table~\ref{tab:amber-hr-sizes}, the inner radius of the wind
launching region is only a  few times larger than the stellar radius
($1.7$--$3.0\,R_{*}$) for MWC~275, HD~98922 and HD~58647, which
have similar spectral types (A1 and B9). On the other hand, it is much larger
($R_{\mathrm{wi}}=17.5\,R_{*}$) for MWC~297 which has a much 
earlier spectral type (B1.5). This may suggest that the environment
of the wind-launching regions in MWC~297 might be different from
those in MWC~275, HD~98922 and HD~58647. The difference between
the wind-launching region of MWC~297 and those of the lower
luminosity stars may be caused by the difference in the strength of 
stellar and/or disc radiation (pressure) which may influence the
wind launching radius and wind dynamics (e.g.~\citealt{Drew:1997};
\citealt*{Proga:1999}).

The physical origin of the relation in 
equation~\ref{eq:dw-inner-radius} is unknown. The increase in
$R_{\mathrm{wi}}$ for the stars with lower luminosities (MWC~275,
HD~98922 and HD~58647) may simply reflect the increase in their
stellar radii (the fourth column in Table~\ref{tab:amber-hr-sizes}),
which, in turn, set the size of magnetospheric radius or the inner
radius of the accretion disc from which the wind arises.  The cause of
the increase in $R_{\mathrm{wi}}$ from the three stars with lower
luminosities (MWC~275, HD~98922 and HD~58647) to MWC~297 is even more
uncertain since the stellar radius of MWC~297 is not significantly
larger than the others.

Another difference between MWC~297 and the other three stars in our
sample can be found in their K-band continuum emission
radii. Fig.~\ref{fig:compare-sizes} also shows the radius of the
K-band continuum-emitting ring for each object along with the expected
dust sublimation radii (assuming the dust sublimation temperature of
1500\,K) computed from the size-luminosity relation in MO05. While the
ring radii of MWC~275, HD~98922 and HD~58647 follow the expected
size-luminosity relation, that of MWC~297 falls below the expected
value for the dust sublimation temperature of 1500~K. The continuum
ring radius is about 10 times larger than the inner radius of the disc
wind in MWC~275, HD~98922 and HD~58647, but it is only 1.1 times
larger for the luminous MWC~297.  The tendency for the high luminosity
Herbig Be stars ($L_{*}>10^{3}\,\Lsun$) to have an `undersized' dust
sublimation radius is pointed out by MO05. They suggest that the
innermost gas accretion disc might be optically thick for luminous
Herbig Be stars, and it partially blocks the stellar radiation on to
the disc, hence causing a smaller dust sublimation radius (see also
\citealt{Eisner:2004}).

The outflow, from which the observed
Br$\gamma$ emission line originates, is most likely formed in
magnetohydrodynamical processes, in either (1)~magneto-centrifugal disc
wind (e.g.~\citealt{blandford:1982,ouyed:1997,krasnopolsky:2003}) or
(2)~a wind launched near the disc-magnetosphere interaction region
(e.g.~an X-wind -- \citealt{shu:1994,Shu:1995} or a conical wind --
\citealt{Romanova:2009}).  The (accretion-powered) stellar wind 
(e.g.~\citealt{decampli:1981, hartmann:1982, Strafella:1998,
  matt:2005, Cranmer:2009}) is unlikely the origin of Br$\gamma$
emission line in the case of HD~58647 because the line computed with such a wind
would have a single-peaked profile while the observed Br$\gamma$ is
double-peaked with its peak separation of $\sim160\,\kmps$
(Fig.~\ref{fig:amber-summary}).  Our radiative transfer model uses a
simple disc wind plus a compact magnetosphere, and we have
demonstrated that this model is able to
reproduce the interferometric observation with the VLTI-AMBER
(Fig.~\ref{fig:amber-model-best}). However, the disc wind in this
model could also resemble the flows found in the conical wind and X-wind
models, in which  the mass ejection region is also   
concentrated just outside of a magnetosphere.  
This is because the inner radius of the disc wind model (Model~F in 
Table~\ref{tab:model-param}) is also located just outside of 
the magnetosphere. Further, the
local mass-loss rate ($\dot{m}$, the mass-loss rate per
unit area) on the accretion disc in our model is a strong
function of $w$ (the distance from the star on the accretion disc
plane), i.e. $\dot{m} \propto w^{-2.4}$. Thus, the mass-loss flux is
 concentrated just outside of the magnetosphere as in the conical
and X-wind models. It would be difficult to exclude the conical or X-wind as a
possible outflow model in this case. Only when the inner radius of the
wind launching region is significantly larger than the corotation
radius of a star (as in MWC~297, Table~\ref{tab:amber-hr-sizes}), they
could be safely excluded.   

Finally, the rate of angular momentum loss by the disc wind
($\dot{J}_{\mathrm{dw}}$) can be directly calculated in our model
(Models~E and F) because we adopted the explicit forms of the
mass-loss rate per unit area and the wind velocity law
(Section~\ref{subsub:DW-config}).  In the context of a
magneto-centrifugal disc wind model, the specific angular momentum
($j$) transported along each magnetic field line can be written as
\begin{equation}
     j = \left(r_{\mathrm{A}}/r_{0}\right)^2 \, j_{0}
     \label{eq:specific-angular-momentum}
\end{equation}
(e.g. \citealt{Pudritz:2007}) where $r_{A}$ and $r_{0}$ are the
Alfv\'{e}n radius and the wind launching radius, respectively, whereas
$j_{0}$ is the specific angular momentum of the disc at $r_{0}$. Here,
we assume $(r_A/r_0) \approx 3$ as in \citet{Pudritz:2007}.  Using the
total wind mass-loss rate found in the Models~E and F
($\dot{M}_{\mathrm{dw}} = 4.5\times10^{-8}\,\MsunPerYear$,
Table~\ref{tab:model-param}) as the normalization constant, we can
integrate the angular momentum carried away by the wind over the whole
wind launching area (from $R_{\mathrm{wi}}=1.7\,R_{*}$ to
$R_{\mathrm{wo}}=23.5\,R_{*}$) to obtain the rate of the total angular
momentum loss by the disc wind ($\dot{J}_{\mathrm{dw}}$). We find this
value as $\dot{J}_{\mathrm{dw}} = 9.9\times10^{38}$\,erg.

On the other hand, the rate of angular momentum loss
    ($\dot{J}_{\mathrm{disc}}$) that is required at the disc wind
outer radius $R_\mathrm{wo}$ to maintain a steady accretion in the
Keplerian disc with a mass-accretion rate $\dot{M}_{\mathrm{a}}$ is
\begin{equation}
      \dot{J}_{\mathrm{disc}} = \dot{M}_{\mathrm{a}}\, j_{\mathrm{wo}}
      = \dot{M}_{\mathrm{a}}\, \left(GM_{*}\,R_{\mathrm{wo}} \right)^{1/2}
      \label{eq:angular-momentum-loss}
\end{equation}
where $j_{\mathrm{wo}}$ is the specific angular momentum of the gas at
the disc wind outer radius $R_{\mathrm{wo}}$
(e.g.~\citealt{Pudritz:1999}).  Using the mass-accretion rate for
HD~58647, $\dot{M}_{\mathrm{a}} = 3.5\times10^{-7}\,\MsunPerYear$
(Table~\ref{tab:stellar-param}), we find $\dot{J}_{\mathrm{disc}} =
1.7\times10^{39}$\,erg.  Hence, the ratio $\dot{J}_{\mathrm{dw}} /
\dot{J}_{\mathrm{disc}}$ is about 0.58. Interestingly, this is very
similar to the ratio found by \citet{Bacciotti:2002} in the
observation of the classical T Tauri star DG~Tau. 

Our model suggests that about 60~per~cent of the angular momentum loss
rate is in the disc wind. This in turn indicates that the disc wind
plays a significant role in the angular momentum transport, and it
dominates the role of the X-wind (if coexists) since the rate of
angular momentum loss by the latter is expected to be much smaller
(e.g.~\citealt{Ferreira:2006}).

%%%%%%%%%%%%%%%%%%%%%%%%%%%%%%%%%%%%%%%%%%%%%%%%%%%%%%%%%
\begin{table*}

\caption{The inner radii of disc wind launching regions
  ($R_{\mathrm{wi}}$) and the continuum-emitting ring
  ($R_{\mathrm{ring}}$) of the Herbig Ae/Be stars observed with the VLTI-AMBER
  in the high spectral resolution mode (including this study). \label{tab:amber-hr-sizes} }

\begin{center}

\begin{center}
\begin{tabular}{ccccccccl}
\hline 
Object & Sp. type & $L_{*}$ & $R_{*}$ & \multicolumn{2}{c}{$R_{\mathrm{wi}}$} & \multicolumn{2}{c}{$R_{\mathrm{ring}}$} & References\tabularnewline
 & $\cdots$ & ($\Lsun$) & ($\Rsun$) & (au) & ($R_{*}$) & (au) & ($R_{*}$) & \tabularnewline
\hline 
MWC~275 & A1 V & $40^{b}$ & $2.3^{a}$ & $0.02^{a}$ & $2.0^{a}$ &
$0.23^{b}$ & $21.5^{b}$ & $^{a}$\citet{GarciaLopez:2015},
$^{b}$\citet{Monnier:2005} \tabularnewline
HD~58647 & B9 IV & $412$ & $6.2$ & $0.05$ & $1.7$ & $0.68$ & $23.5$ & This work\tabularnewline
HD~98922 & B9 V & $640$ & $7.6$ & $0.1$ & $3.0$ & $0.7$ & $19.8$ & \citet{Caratti:2015}\tabularnewline
MWC~297 & B1.5V & $10600$ & $6.1$ & $0.5$ & $17.5$ & $0.56$ & $19.7$ & \citet{Weigelt:2011}\tabularnewline
\hline 
\end{tabular}
\par\end{center}

\end{center}

\end{table*}
%%%%%%%%%%%%%%%%%%%%%%%%%%%%%%%%%%%%%%%%%%%%%%%%%%%%%%%%%

%%%%%%%%%%%%%%%%%%%%%%%%%%%%%%%%%%%%%%%%%%%%%%%%%%%%%%%%%
%%%%%%%%%%%%%%%%%%%%%%%%%%%%%%%%%%%%%%%%%%%%%%%%%%%%%%%%%

\subsection{Is there a magnetosphere in HD~58647?}
\label{sub:discuss-ma}

The presence of a magnetosphere in HD~58647 is suggested by a variety
of observations. For example, the spectroscopic study of HD~58647 by
\citet{Mendigutia:2011} shows a hint for an inverse P-Cygni profile
or a redshifted absorption component in \ion{He}{i}~5876~\AA{}, which
might be formed in a magnetospheric accretion funnel. Interestingly, a
recent spectroscopic survey of Herbig Ae/Be stars by
\citet{Cauley:2015} has shown that about 14~per~cent of their samples
exhibit the inverse P-Cygni profile in \ion{He}{i}~5876~\AA{} (see
also \citealt{Cauley:2014}). Further, the spectropolarimetic (linear)
observation of H$\alpha$ in HD~58647 by \citet{Vink:2002}, \citet{Mottram:2007}
and \citet{Harrington:2009} show a large polarization change in the
line and the pattern on the Stokes parameter $QU$ plane is a loop ---
indicating that a part of H$\alpha$ emission is intrinsically
polarized and the emission is scattered by the gas with a flattened
geometry.  A similar pattern on the $QU$ plane are also seen in the
classical T Tauri stars and Herbig Ae stars.  \citet{Mottram:2007}
suggested that the intrinsically polarized H$\alpha$ (at least
partially) originates from a magnetosphere.  Lastly, a recent
spectropolarimetric observation by \citet{Hubrig:2013} showed
that the star has a mean longitudinal magnetic field with strength
<$B_{z}$>=$218\pm69$~G, supporting a possible presence of a
magnetosphere in HD~58647.

Based on these evidences, a magnetosphere with a small radius was
introduced in the radiative transfer models for the Br$\gamma$
line profile and the interferometric data
(Sections~\ref{sub:model-line-prof} and \ref{sub:model-vis}). In
Sections~\ref{sub:model-line-prof}, we found that the small
magnetosphere does not significantly contribute to the Br$\gamma$ line
emission (Fig.~\ref{fig:ma-tm-effect}).  On the other hand, we find
the disc wind is the main contributor for the Br$\gamma$ emission in
HD~58647 (Fig.~\ref{fig:compare-ma-dw}). Since the mass-accretion rate
of HD~58647 is relatively high ($\dot{M}_{\mathrm{a}}=3.5\times
10^{-7}\,\MsunPerYear$, \citealt{Brittain:2007}), it may contribute a
non-negligible amount of continuum flux in the K-band. For example, in
Model~F, approximately 10~per~cent of the total continuum flux is from the
magnetosphere. This has at least two consequences: (1)~the normalized
line profile will be slightly weaker because of the larger total
continuum flux (Fig.~\ref{fig:compare-ma-dw}), and (2)~the continuum
visibilities would be larger since the magnetosphere is 
unresolved (its continuum emission is compact)
(Fig.~\ref{fig:amber-model-best}).  In other words, we might
slightly underestimate e.g. the disc wind mass-loss rate if a magnetosphere
is not included in the model since it would produce a slightly stronger line
(due to a weaker continuum). We might also slightly underestimate 
the size for the continuum-emitting ring when the magnetosphere is not
included in the model. A slightly larger ring size is needed to balance
the increase in the visibility due to the emission from the magnetosphere.

One can estimate the dipolar magnetic field strength of HD\,58647 at the magnetic 
equator ($B_{*}$) using the standard relation between the
dipole magnetic field strength and stellar parameters found in
\citet{Koenigl:1991} and \citet*{Johns-Krull:1999}, i.e.
\begin{eqnarray}
B_{*} & = & K
            \left(\frac{\epsilon}{0.35}\right)^{7/6}
            \left(\frac{\beta}{0.5}\right)^{-7/4}
            \left(\frac{M_{*}}{\Msun}\right)^{5/6} \\
      &   & \times\left(\frac{\dot{M}_{\mathrm{a}}}{10^{-7}\,\MsunPerYear}\right)^{1/2}
            \left(\frac{R_{*}}{\Rsun}\right)^{-3}
            \left(\frac{P_{*}}{1.0\,\mathrm{d}}\right)^{7/6}
\label{eq:konigl}
\end{eqnarray}
where $K$ is a constant: $3.43\times10^{3}\,\mathrm{G}$. The
parameters $\beta$ and $\epsilon$ represent the efficiency of the
coupling between the stellar magnetic field and the inner regions of
the disc, and the ratio of the stellar angular velocity to the
Keplerian angular velocity at the inner radius of a magnetosphere,
respectively. Here, we adopt $\beta=0.5$ and $\epsilon=0.35$, as in
\citet{Koenigl:1991} and \citet{Johns-Krull:1999}. By using the
stellar parameters used in our model (Table~\ref{tab:stellar-param})
in equation~\ref{eq:konigl}, we find $B_{*} = 240$\,G. This is very
similar to the mean longitudinal magnetic field strength
<$B_{z}$>=$218\pm69$~G found in the spectropolarimetric
observations of HD~58647 by \citet{Hubrig:2013}.

\subsection{A Keplerian disc as a main source of Br$\gamma$ emission?}
\label{sub:discuss-kdisc}

So far, we have argued that the double-peaked Br$\gamma$ emission line
profile seen in HD~58647 is mainly formed in a disc wind, which
is rotating near Keplerian velocity at the base of the wind. The model
with a disc wind and a magnetosphere (Model~F in
Fig.~\ref{fig:amber-model-best}) was also able to reproduce the
interferometric observations by VLTI-AMBER. 
However, the double-peaked Br$\gamma$ line profile could be also
formed in a gaseous Keplerian disc (without a disc wind), as often
demonstrated in the models for (more evolved) classical Be stars
(e.g.\,\citealt{Carciofi:2008, Kraus:2012c, Meilland:2012,
  Rivinius:2013}). In addition, we find 
that the emission maps computed for different velocity channels across
Br$\gamma$ shown in Fig.~\ref{fig:vel-image} is rather similar to those
computed with a simple Keplerian disc (e.g.\,\citealt{Kraus:2012c}; \citealt{Mendigutia:2015}). 
The similarity naturally occurs because the Br$\gamma$ emission in our
disc wind model arises mainly from near the base of the wind where the
Keplerian rotation of the wind is dominating over the poloidal motion.
For this reason, it would be difficult to exclude a simple Keplerian disc
as a main source of the Br$\gamma$ emission in HD~58647 based
solely on our VLTI-AMBER observations.  

\begin{figure}

\begin{center}

  \includegraphics[clip,width=0.45\textwidth]{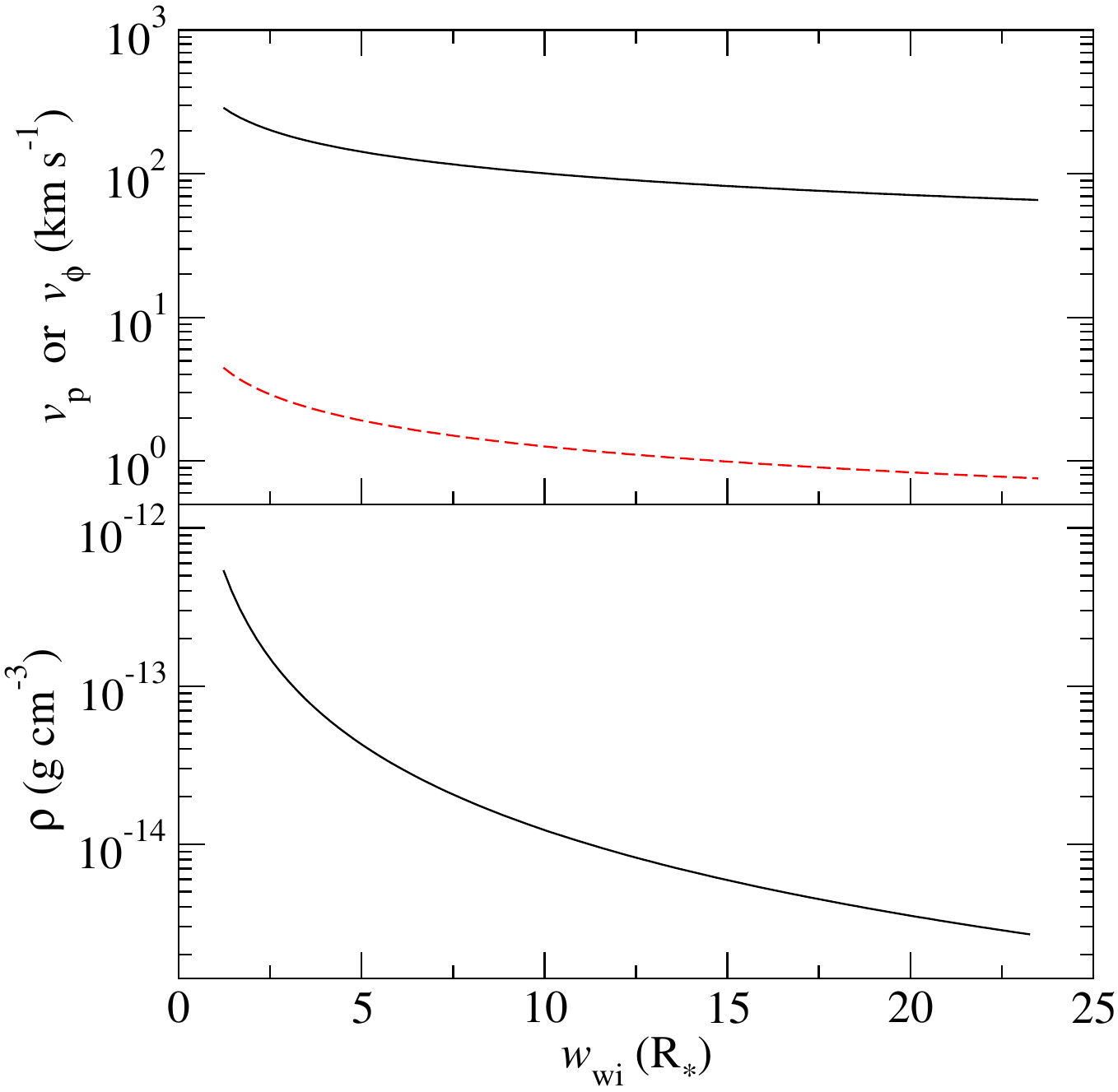}

\caption{ A comparison of the velocity components (upper panel) and
  density (lower panel) at the base of the disc wind model used in
  Model~F.  The poloidal (dash) and Keplerian (solid) velocity
  components ($v_{\mathrm{p}}$ and $v_{\phi}$, respectively) are
  plotted as a function of the distance ($w_{i}$) from the
  symmetric/rotation axis.  The range of $w_{i}$ is from
  $R_{\mathrm{wi}}=1.7\,R_{*}$ to $R_{\mathrm{wo}}=23.5\,R_{*}$ as in
  Model~F (Table~\ref{tab:model-param}).  The poloidal speed of the
  disc wind at the base is calculated by setting $l=0$ in
  equation~\ref{eq:discwind-poloidal-velocity} in
  Section~\ref{subsub:DW-config}. The corresponding density ($\rho$)
  of the disc wind is computed using equation~6 in
  \citet{Kurosawa:2006}.  The Keplerian component dominates the
  poloidal component at all the wind launching radii. The wind density
  at the base decreases from $\sim5\times10^{-13}$ to
  $\sim3\times10^{-15}\,\mathrm{g\,cm^{-3}}$ as the wind launching
  radius increases.
  \label{fig:rho-vel-base} 
} 

\end{center}

\end{figure}

Fig.~\ref{fig:rho-vel-base} shows the poloidal and azimuthal velocity
components of the disc wind at its base for Model~F, plotted as a
function of the distance from the symmetric/rotation axis. The
corresponding density at the base of the disc wind is also shown in
the same figure as a reference.  In our model, the poloidal speed of
the disc wind ($v_{\mathrm{p}}$) is set by
equation~\ref{eq:discwind-poloidal-velocity} in
Section~\ref{subsub:DW-config}. The poloidal speed of the wind at its
base can be found by setting $l=0$ in the equation, i.e.
$v_{\mathrm{p}}(w_{i}, l=0) = c_{\mathrm{s}}(w_{i})$ where
$c_{\mathrm{s}}(w_{i})$ the local sound speed, and $w_{i}$ is the
distance from the symmetry/rotation axis. Hence, the poloidal speed at
the base of the wind is equal to the local sound speed. The sound
speed at the inner disc wind launching radius
($R_{\mathrm{wi}}=1.7\,R_{*}$) is about 4\,$\kmps$ (assuming
$T\sim2000$\,K as in \citealt{Muzerolle:2004}).  This is significantly
smaller than the (wind) Keplerian speed at the same location,
i.e. 289\,$\kmps$. Although not shown here, as a simple test, we
computed Br$\gamma$ line profiles using exactly the same density
distribution as in Model F (magnetosphere + disc wind), but with
$v_{\mathrm{p}}$ manually boosted by a constant factor to examine how
large $v_{\mathrm{p}}$ should be for the Br$\gamma$ line profile to
show a line asymmetry or a deviation from the Keplerian velocity
dominated Br$\gamma$ line profile seen in Model~F. In this experiment,
we found that the Br$\gamma$ line profile started to show the effect
of $v_{\mathrm{p}}$ when it was boosted by a factor of $\sim 5$. The
corresponding $v_{\mathrm{p}}$ at the base of the wind at the inner
wind radius ($R_{\mathrm{wi}}$) is about 25\,$\kmps$.  In the original
disc wind velocity distribution (without the boosting factor),
$v_{\mathrm{p}}$ can reach 25\,$\kmps$ at about 2\,$R_{*}$ away from
the wind launching point along the streamline emerging from the inner
disc wind radius ($R_{\mathrm{wi}}$). If the Br$\gamma$ emission
mainly arose from this region or a location farther away from the
launching point, we would be able to see the effect of the poloidal
velocity. However, this is not the case for our model (Model~F),
i.e. the Br$\gamma$ emission seems to originate much closer to the
disc surface.

As seen in Section~\ref{subsub:line-DW-model}, to produce a Br$\gamma$
emission line with its strength comparable to the one seen in the
observation, the temperature of the gas must be rather high ($\sim
10^{4}$\,K) and the emission volume must be sufficiently large. In the
study of thermal properties of Keplerian discs (without a disc wind)
around classical Be star, \citet{Carciofi:2006} showed that the gas
temperature in the upper layers of discs could reach $\sim 10^{4}$\,K,
although they considered stars with $T_\mathrm{eff}=19000$\,K which is
much higher than that for HD~58647 ($T_\mathrm{eff}=10500$\,K,
Table~\ref{tab:stellar-param}). It is uncertain if the same
model is applicable to HD~58647 and if it can produce gas with high
enough temperature ($\sim 10^{4}$\,K) using the stellar parameters of
HD~58647. Furthermore, in the study of the thermal property of gaseous
discs irradiated by Herbig Ae/Be stars,
\citet{Muzerolle:2004} found the gas temperatures in the upper
layers of accretion discs were relatively low ($\sim$2000\,K). This
is certainly too low for the formation of a Br$\gamma$ emission line. 
On the other hand, the disc wind gas has favorable conditions for Br$\gamma$
emission because of the acceleration and heating of the gas in
the disc wind by e.g.~ambipolar diffusion (e.g.~\citealt{Safier:1993};
\citealt{Garcia:2001}) and internal jet shocks (e.g.~\citealt{Staff:2010}),
which can bring the temperature to $\sim 10^{4}$\,K.    

To compute the emission line profiles from a Keplerian disc
with self-consistent density and temperature structures (by
considering hydrostatic and radiative equilibriums), we would need a
major change in our model. This is beyond the scope of this paper;
however, it should be investigated in the future in detail. Only 
then, we would be able to address the issue of distinguishing
interferometric signatures of a disc wind from those of a simple
Keplerian disc. 

%%%%%%%%%%%%%%%%%%%%%%%%%%%%%%%%%%%%%%%%%%%%%%%%%%%%%%%%%
%%%%%%%%%%%%%%%%%%%%%%%%%%%%%%%%%%%%%%%%%%%%%%%%%%%%%%%%%

\section{Conclusions}
\label{sec:conclusions}

We have presented a study of the wind-launching region of the Herbig
Be star HD~58647 using high angular and high spectral ($R=12000$) resolution
interferometric observations of the Br$\gamma$ line-emitting region.  The
high spectral resolution of AMBER provides us many spectral channels across the line to
investigate the wind origin and properties of HD~58647
(Fig.~\ref{fig:amber-summary}).  The star displays
double peaks in Br$\gamma$ and in the wavelength-dependent visibility
curves. The differential phase curves show S-shaped variations around
the line centre. 

Based on a simple geometrical ring fit for the continuum
visibilities (Fig.~\ref{fig:ring-fit}), we find the K-band
continuum-emitting ring size is around 2.0~mas, which corresponds to
0.64~au at the distance of 318~pc (VL07). The visibility level
increases in the line (by $\sim 0.1$) for the longest project baseline
(88\,m), indicating that the extent of the Br$\gamma$ emission region
is smaller than the size of the continuum-emitting region, which is
expected to arise near the 
dust sublimation radius of the accretion disc. 

The interferometric data have been analysed using radiative transfer
models to study the geometry, size and physical properties of the
wind.  The double-peaked line profile and the pattern of the
differential phase curves (S-shaped) suggest that the line-emitting
gas is rotating.  We find that a model with a small magnetosphere with
outer radius $R_{\mathrm{mo}}=1.7\,R_{*}$ plus a disc wind with the
inner radius located just outside of the magnetosphere and outer
radius $R_{\mathrm{wo}}=22.5\,R_{*}$ (Model~F) can fit the observed
Br$\gamma$ profile, wavelength-dependent visibilities, differential
and closure phases, simultaneously
(Fig.~\ref{fig:amber-model-best}). The model has shown that the radial
extent of the Br$\gamma$ emission in the disc wind is about 0.5~au
while the inner radius of the K-band continuum-emitting ring is about
0.68~au (Fig.~\ref{fig:vel-image}).  The slightly larger size of the
continuum ring emission, compared to the one obtained in the simple
geometrical model (Section~\ref{sec:ring-model-fit}), is found in the
radiative transfer model because it includes the continuum emission
from the compact magnetosphere ($\sim$\,10~per~cent of the total
K-band flux).  The mass-accretion and mass-loss rates adopted for the
model are $3.5\times10^{-7}$ and $4.5\times10^{-8}\,\MsunPerYear$,
respectively ($\dot{M}_{\mathrm{dw}}/\dot{M}_{\mathrm{a}}=0.13$).  
Consequently, about 60~per~cent of the angular momentum loss
rate required for a steady accretion with the measured
accretion rate is provide by the disc wind
(Section~\ref{sub:discuss-line-region}). 

We find that the small magnetosphere does not
contribute significantly to the Br$\gamma$ line emission significantly
(Fig.~\ref{fig:compare-ma-dw}).  The size of the innermost radius in
the disc wind model (Model F) was compared with those from the earlier
studies Herbig Ae/Be stars with high spectral resolution VLTI-AMBER
observations (\citealt{Weigelt:2011, GarciaLopez:2015,Caratti:2015}).
We found a trend that the inner radius of the wind-launching region
increases as the luminosity of a star increases
(Fig.~\ref{fig:compare-sizes}). 

While the model with a disc wind plus magnetosphere (Model~F) showed
the best fit to the interferometric observations
(Fig.~\ref{fig:amber-model-best}), the disc wind only model (Model E)
could be improved to fit the observations (in particular the
visibility levels for the longest baseline) by using a smaller
continuum ring radius and by adjusting other model parameters
accordingly. Hence, the model proposed here (Model~F) is not necessary
a unique fit to the observations.

As briefly mentioned in Section~\ref{sub:discuss-line-region}, the
AMBER observation could be also explained by the conical wind
(e.g.~\citealt{Romanova:2009}) or X-wind (e.g.~\citealt{shu:1994})
models because the disc wind model that fits the AMBER observation has
an inner radius comparable to the size of the magnetosphere, and its
mass-loss is also concentrated near the disc-magnetosphere interaction
region, as in the conical and X-wind models.
However, the X-wind model alone might have a difficulty in explaining
the rate of the angular momentum loss required for a steady accretion
in the Keplerian disc (Section~\ref{sub:discuss-line-region}).
Also, as mentioned in Section~\ref{sub:discuss-kdisc}, we still
cannot exclude a simple Keplerian disc model as 
a main source of the Br$\gamma$ emission in HD~58647 since the model
could produce a double-peaked Br$\gamma$ emission line and could
produce similar velocity-dependent emission maps to those computed
with a disc wind (Fig~\ref{fig:vel-image}). 

In the future, we need a systematic study to find how the
interferometric signatures differ in different outflow models, perhaps
by using the results of MHD simulations.  
To improve our understanding of the wind properties
(e.g.~temperature), it would be useful to combine the interferometric
observations with high-resolution spectroscopic observations. 
Simultaneous fitting of multiple emission lines in the spectroscopic
observations would provide us tighter constrains on the physical
properties of the gas in the wind.  To improve the interferometric data,
we would need higher signal-to-noise data under good
seeing condition and a larger coverage on the $uv$ plane.

\section*{Acknowledgement}

We thank the anonymous referee who provided us insightful comments and
suggestions which helped improving the manuscript.
We thank the ESO support astronomer Dr.~W.\,J.\,M.~de Wit for his
assistance during our observation. We also thank Dr.~Ignacio
Mendigut{\'{\i}}a for the comments on the optical spectra.  We also thank
Prof.~Makoto Kishimoto and Dr.~Florentin Millour for providing us
their observing tools.  A.K. acknowledges support from the STFC
Rutherford Grant (ST/K003445/1). S.K. acknowledges support from a STFC
Rutherford fellowship (ST/J004030/1). This publication makes
use of VOSA, developed under the Spanish Virtual Observatory project
supported from the Spanish MICINN through grant AyA2011-24052. This
research has made use of the Jean-Marie Mariotti Center \texttt{Aspro}
service.\footnote{Available at http://www.jmmc.fr/aspro} 

\bibliographystyle{mn}
\bibliography{reference}

% Don't change these lines 
\bsp % typesetting comment 
\label{lastpage} 
\end{document}